\documentclass[a4paper,11pt]{article}
\pdfoutput=1 

\usepackage{jinstpub} 

\usepackage{graphicx}
\usepackage{subcaption}
\usepackage{amsmath}
\usepackage{txfonts}
\usepackage{upgreek}
\usepackage{lineno}

\usepackage{appendix}

\newcommand{\nech}{\ensuremath{\mathrm{Ne/(5\%CH_4)~}}}
\newcommand{\arch}{\ensuremath{\mathrm{Ar/(5\%CH_4)~}}}

\newcommand{\dvdrift}{\ensuremath{\mathrm{\Delta V_{drift}}}}
\newcommand{\dvrpwell}{\ensuremath{\mathrm{\Delta V_{RPWELL}}}}

\title{On the localization properties of an RPWELL gas-avalanche detector}

\author[a,1]{L. Moleri\note{Corresponding author.},}
\author[a]{P. Bhattacharya,}
\author[a]{A.E.C. Coimbra,}
\author[a]{A. Breskin,}
\author[a]{and S. Bressler}

\affiliation[a]{Department of Particle Physics and Astrophysics, Weizmann Institute of science,\\
7610001 Rehovot, Israel\\}

\emailAdd{luca.moleri@weizmann.ac.il}

\abstract{A study of the localization properties of a single-element Resistive Plate WELL (RPWELL) detector is presented. The detector comprises of a single-sided THick Gaseous Electron Multiplier (THGEM) coupled to a segmented readout anode through a doped silicate-glass plate of 10$^{10}$~$\Omega\cdot$cm bulk resistivity. Operated in ambient \nech gas, the detector has been investigated with 150~GeV muons at CERN-SPS. Signals induced through the resistive plate on anode readout strips were recorded with APV25/SRS electronics. The experimental results are compared with that of Monte Carlo simulations. The effects of various physics phenomena on the position resolution are discussed. The measured position resolution in the present configuration is 0.28~mm RMS - compatible with the holes-pattern of the multiplier. Possible ways for improving the detector position resolution are suggested.}


\begin{document}
\maketitle
\flushbottom

\section{Introduction}
\label{sec: Introduction}
The Resistive Plate WELL (RPWELL)~\cite{rubin2013first} is a single-element gas-avalanche detector, comprising of a single-sided THick Gaseous Electron Multiplier (THGEM)~\cite{chechik2004thick,breskin2009concise} coupled to a segmented readout anode through a highly-resistive plate. Similarly to the Resistive-Plate Chamber (RPC)~\cite{santonico1981development}, RWELL~\cite{arazi2014laboratory}, Micromegas~\cite{alexopoulos2011spark}, and others, the resistive material was introduced to protect the detector from the occurrence of occasional discharges. With its mm-scale holes pattern, the RPWELL concept is suitable for applications requiring large area-coverage and moderate localization capabilities. Recent experimental results obtained with an RPWELL detector, in the context of digital hadron-calorimetry (DHCAL), are detailed in~\cite{bressler2016first,moleri2016resistive,moleri2016beam}; in these works, the RPWELL detector operated in a discharge-free mode at high gain over a broad dynamic range.  
The present work focuses on the study of the position resolution of a glass-RPWELL detector, with a THGEM coupled to a one-dimensional strips array anode, through a doped silicate-glass resistive plate of 10$^{10}$~$\Omega\cdot$cm bulk resistivity~\cite{wang2010development}.  The goal was to measure the position resolution and to understand the physics phenomena governing it - as to allow for designing detectors with optimized localization capabilities. Previous works investigating position resolution of THGEM-based detectors are reported in~\cite{cortesi2007investigations,cortesi2009thgem,silva2013x,lopes2013position}; they dealt with localization studies of soft x-rays and UV photons, using standard image-analysis techniques and demonstrated a position resolution ranging from 0.3~mm to 2.3~mm FWHM depending on the specific detector configuration, operation gas and type of radiation.  In this work, we investigated the detector with relativistic muons; each muon track was referenced to a high-resolution tracker, and the detector hits were analyzed on an event-by-event basis using a procedure similar to the one presented in~\cite{abusleme2016performance} for a Thin-Gap Chamber detector. This method allows studying the local differences in position resolution due to the detector geometry. 
Similar studies conducted on different detectors, like resistive micro-WELL~\cite{lener2016mu}, with its $\sim$7-fold smaller pitch in respect to the RPWELL, or Resistive-Plate Chamber with its continuous sensitive area~\cite{aielli2014rpc}, yielded RMS resolutions down to 52~$\upmu$m and 70~$\upmu$m (estimated) respectively. The present results show that in the RPWELL the localization capability is limited to significantly higher values mainly by the THGEM-holes pattern; possible improvements are suggested.
In section 2 we present the experimental setup and methodologies, followed by test-beam results in section 3. A detailed comparison with Monte Carlo simulations is described in section 4, followed by a discussion in section 5 and an appendix on the simulation method.

\begin{figure}[h]
\begin{subfigure}[t]{0.6\textwidth}\caption{}
\includegraphics[scale=0.15]{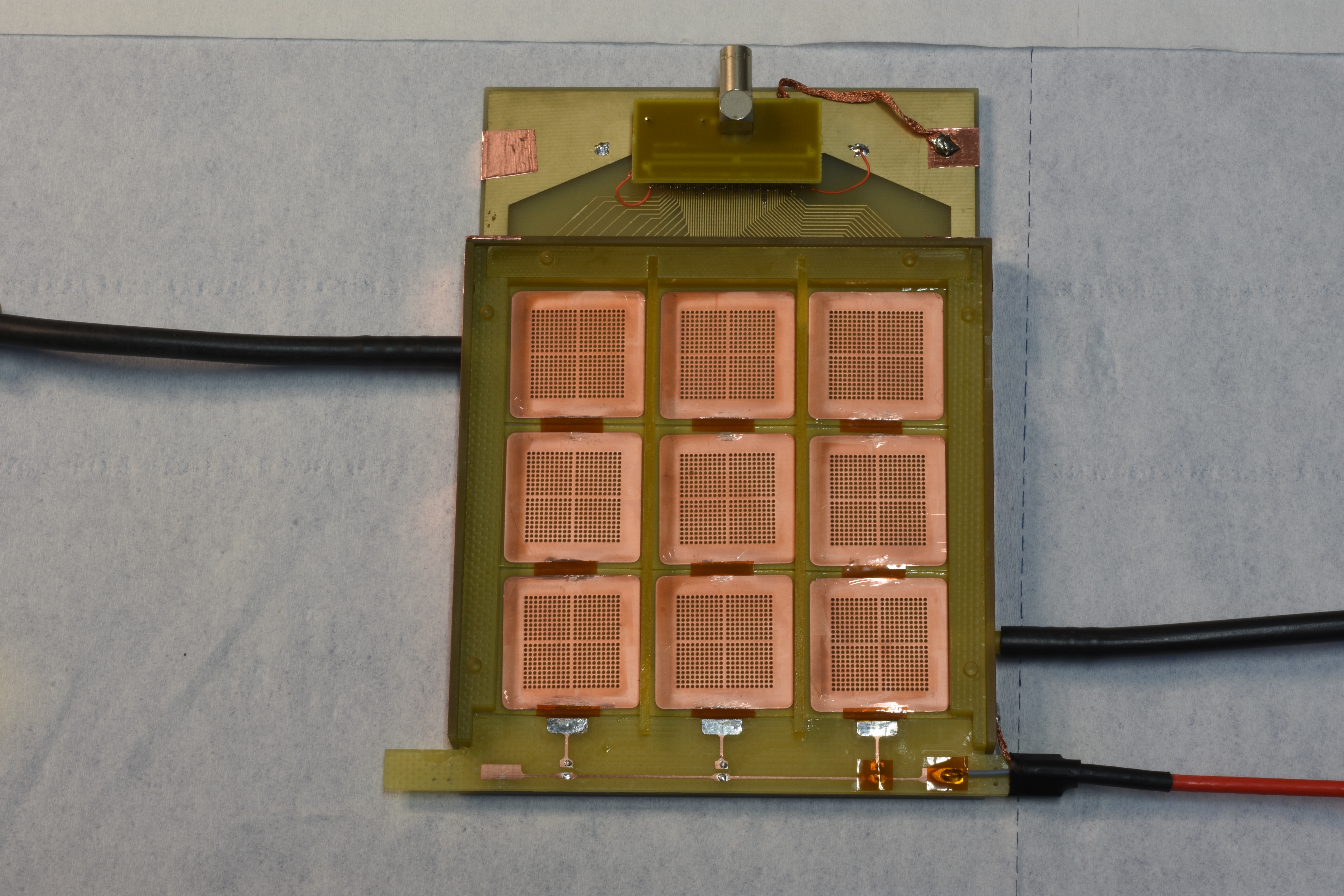}
\end{subfigure}
\begin{subfigure}[t]{0.4\textwidth}\caption{}
\includegraphics[scale=1]{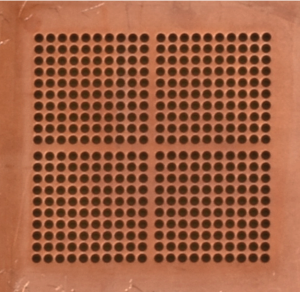}
\end{subfigure}
\caption{A photograph of the RPWELL detector's THGEM-electrodes plane (cathode removed), made of 9 tiles (a) with a 0.96~mm pitch square holes-pattern (b). Note the larger 1.3~mm pitch between the central rows.}\label{fig: RPWELL detector}
\end{figure}

\section{Experimental setup and methodology}
\label{sec: setup and methodology}

\subsection{The RPWELL detector}\label{sec: RPWELL}
\begin{figure}[h]
\centering
\includegraphics[scale=1]{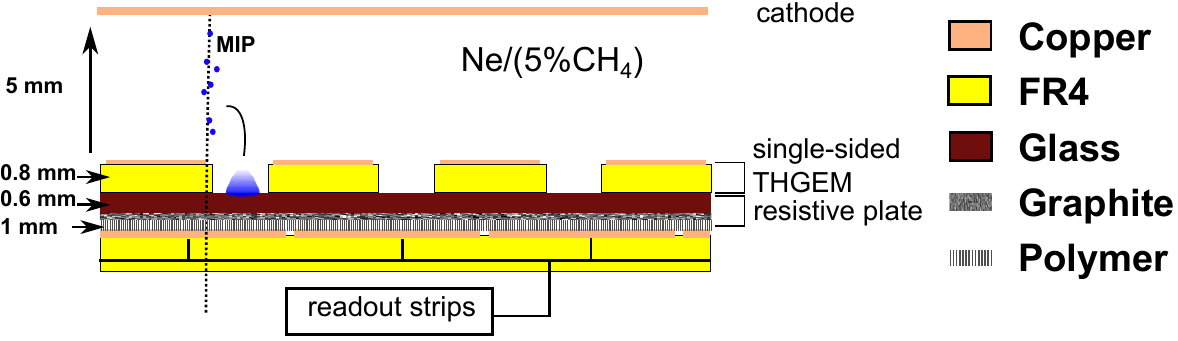}
\caption{A schematic (not to scale) view of the RPWELL detector assembly and operation. From top to bottom the elements are: cathode, single-sided THGEM, resistive plate and readout strips.}\label{fig: RPWELL scheme}
\end{figure}

We assembled an RPWELL detector with an area of 9$\times$9~cm$^2$; the multiplier is segmented into nine 3$\times$3~cm$^2$ THGEM-electrode tiles as shown in figure~\ref{fig: RPWELL detector}-a. Each 0.8~mm thick THGEM tile had a 2$\times$2~cm$^2$ squared hole pattern (figure~\ref{fig: RPWELL detector}-b): 0.5~mm hole-diameter, with 0.1~mm etched rims and a pitch of 0.96~mm (except for a central cross with a 1.3~mm pitch). This peculiar geometry permitted comparing present results with that of~\cite{bressler2016first}. The detector scheme and operation principle is shown in figure~\ref{fig: RPWELL scheme}.  Particle-induced ionization electrons deposited within the 5~mm long drift gap (defined by the cathode and the top THGEM electrode) are focused into the THGEM holes, where they initiate an avalanche multiplication. A resistive plate (RP) is directly coupled to the bare THGEM-bottom face; the avalanche induces a charge through the RP, onto the strips-patterned anode.  The RP in this work is made of 0.6~mm thick doped silicate glass with a bulk resistivity of $\sim$10$^{10}~\Omega\cdot$~cm~\cite{wang2010development}. The avalanche electrons travel across the RP and are evacuated to ground through the RP bottom surface, coated with a thin layer of graphite (similarly to the RWELL~\cite{arazi2014laboratory}); its surface resistivity is $\sim$3~M$\Omega/\Box$ and it is connected to ground through a side copper strip. The RP bottom side is coated with a 1~mm thick polymer epoxy layer, on top of the resistive film, to provide mechanical support to the glass and to protect the resistive layer when handling the detector.  In this configuration the RP is electrically decoupled from the readout anode (placed at 1.6~mm from the THGEM bottom, 1~mm from the resistive film), which had copper strips, divided into three groups of 1~mm, 1.5~mm and 2~mm pitch, separated from each other by 50~$\upmu$m. Each group of strips recorded signals induced from three THGEM tiles. Signals from the strips were processed by an APV25/SRS system~\cite{martoiu2013development,french2001design} and the data were analyzed with dedicated software, as described below.
The detector was operated in \nech at ambient conditions, at a flow of 30~cc/min. The electrodes were polarized with individual HV power-supply CAEN A1833P and A1821N boards, remotely controlled with a CAEN SY2527 unit. The voltage and current of each channel were monitored and stored. All inputs were connected through low-pass filters.  Unless stated otherwise, the voltage across the drift gap was kept at \dvdrift = 250~V, resulting in a drift field of 0.5~kV/cm. The voltage across the RPWELL, \dvrpwell,  varied between 900-975~V.

\subsection{Tracking, readout system and analysis framework}
\label{sec: tracking and readout}

The detector was installed at the CERN-SPS H2 test-beam area and investigated with a flux of $\sim$50~Hz/cm$^2$ 150~GeV muons. The CERN-RD51 Micromegas telescope provided the trigger and precise tracking of the muons. A detailed description of the telescope is given in~\cite{karakostas2010micromegas,karakostas2012telescope}. The RPWELL chamber was placed along the beam line in between the layers of the telescope. The data acquisition system, common to the RPWELL and the tracker, was based on the SRS/APV25 readout electronics~\cite{martoiu2013development,french2001design}). The data processing framework is described in detail in~\cite{bressler2016first}. For each readout channel a threshold relative to the pedestal noise was set, using a common Zero-order Suppression Factor (ZSF).  

\begin{figure}[h]
\centering
\includegraphics[scale=0.3]{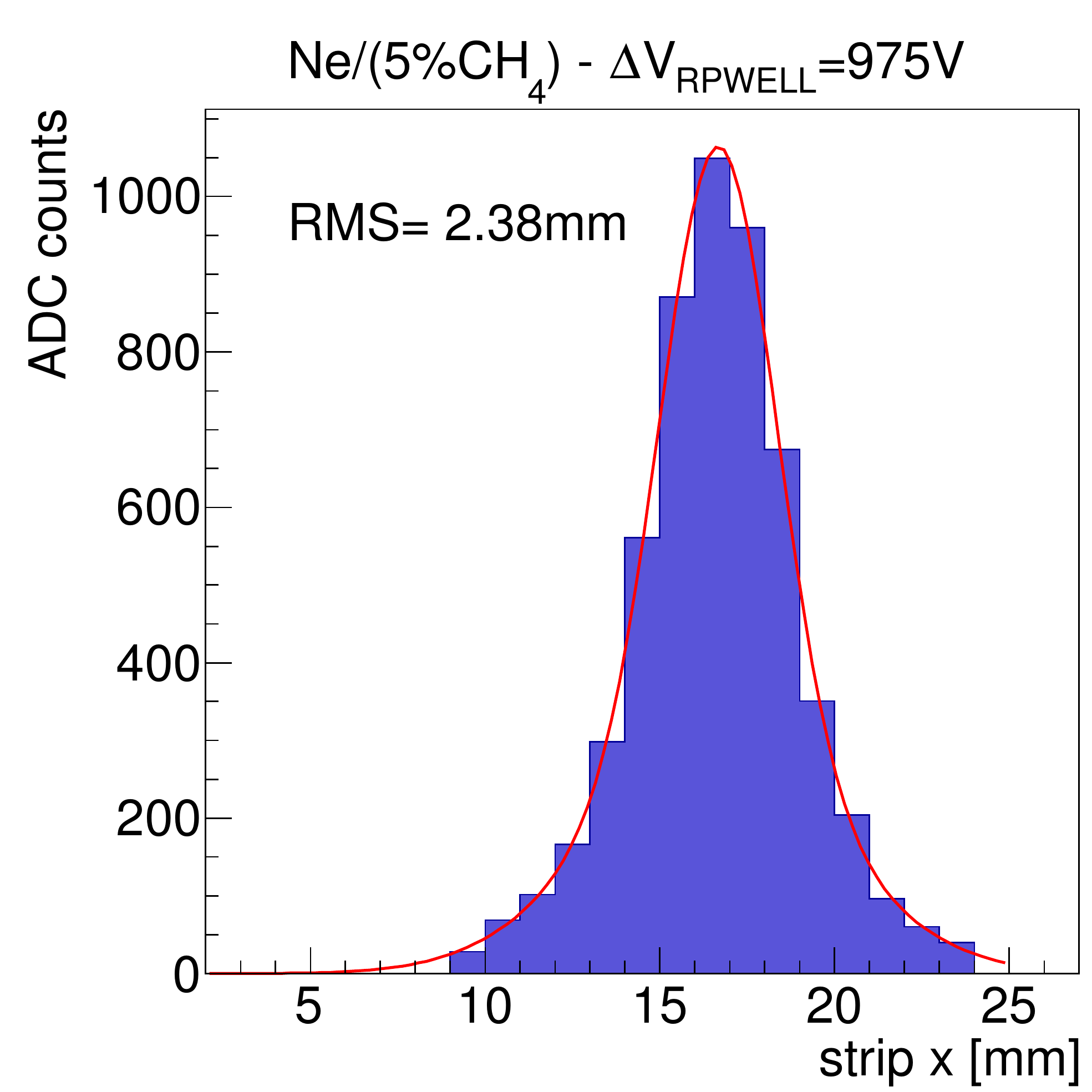}
\caption{The typical induced-charge distribution, at normal incidence, on the 1~mm readout strips measured by the SRS/APV25 electronics. Fit to a double-Gaussian yields a RMS-value of 2.38~mm.}\label{fig: induced signal}
\end{figure}

The tracker and the RPWELL were aligned using the data measured in dedicated runs. We define the THGEM surface as an x-y plane, with the x-axis being perpendicular to the readout strips; this 1-D detector-readout did not provide any information in the y-axis.  For each strip-pitch region, only tracks hitting the detector in the central part of a tile, 13~mm along the x-axis, were considered; this assured that the induced signals are confined within the area covered by the strips.
Among all the muons measured in the RPWELL, we selected for analysis only those that yielded a reconstructed track perpendicular to the x-y plane (or at a fixed angle in dedicated measurements). 
In figure~\ref{fig: induced signal}, we show the typical induced-charge distribution on the 1~mm readout strips, measured by the SRS/APV25 electronics. The $\sim$2.4~mm RMS value of a double-Gaussian fit is independent of the total charge. The lateral spread of the induced-charge distribution depends, instead, strongly on the maximum distance between the readout plane and the fast moving charges produced in the avalanche, i.e. from the resistive-plate top face (as shown also in~\cite{cortesi2007investigations}). This distance was chosen large enough to distribute the induced signal among several strips, allowing defining the center-of-gravity of the induced-charge distribution with a resolution superior to the strip pitch. Considering charge clusters of neighboring strips with signal above threshold, we define a cluster position: x$\mathrm{_{cl}}$ = $\mathrm{\sum_i(x_i\cdot q_i)/\sum_iq_i}$, where x$\mathrm{_i}$  is the position of the center of strip \textit{i} and q$\mathrm{_i}$ is the charge induced on it. The cluster charge is defined as Q$\mathrm{_{cl}}$ = $\mathrm{\sum_iq_i}$. The strip multiplicity is defined as the number of strips in a cluster.  For each event, we calculate the residual along the x-axis, defined as RES = x$\mathrm{_{tr}}$-x$\mathrm{_{cl}}$ , where x$\mathrm{_{tr}}$ is the reconstructed track intersection with the detector. We define the detector position resolution as the RMS of the Gaussian fit of the residuals histogram. The contribution of the position resolution of the telescope ($\sim$0.05~mm RMS) has a negligible effect on that of the RPWELL, and is not taken into account in this work. A matching parameter, W~[mm], is defined as the maximal distance allowed between the reconstructed cluster position and the intersection point of the track with the RPWELL. W-values of the order of 3~mm, significantly larger than the RPWELL resolution, were used to avoid biasing the measured position resolution (see below).
The effective gain of the detector is estimated as follows: a MIP traversing through the 5~mm drift gap of \nech produces on average 20 electron-ion pairs~\cite{sauli2014gaseous}. About half reach the THGEM within 100~ns~\cite{peisert1984drift}, which is approximately the shaping time of the APV25 chip~\cite{french2001design}. Hence, the detector gain is estimated by the total measured charge divided by the charge of 10 primary electrons. We emphasize that the shaping time of the APV25 is much shorter than the rise time of the RPWELL ($\sim$2~$\upmu$s). Hence, the estimated effective gain is a factor of $\sim$3 lower than the total gas gain~\cite{rubin2013first}. 
The detector performance is measured for different ZSF values in the range between 0.7 and 5. Over this range, the strip multiplicity measured with the 1~mm strips varied between 11 to 7 respectively and the position resolution is affected by $\sim$2$\%$. We conclude that within this range of ZSF values, the strip multiplicity is sufficiently high. Over this work we chose ZSF= 1, which we found appropriate for excluding most of the noise hits. In these conditions, the strip multiplicity with the 1.5~mm strips is 6, also not affecting the position resolution. The 2~mm strips were not used in this study. Varying the matching parameter W from 2 to 5 mm did not affect the results; its value is fixed to 3~mm. 

\section{Test-beam results}
\label{sec: Results}

\begin{figure}[h]
\begin{subfigure}[t]{0.5\textwidth}\caption{}
\includegraphics[scale=0.3]{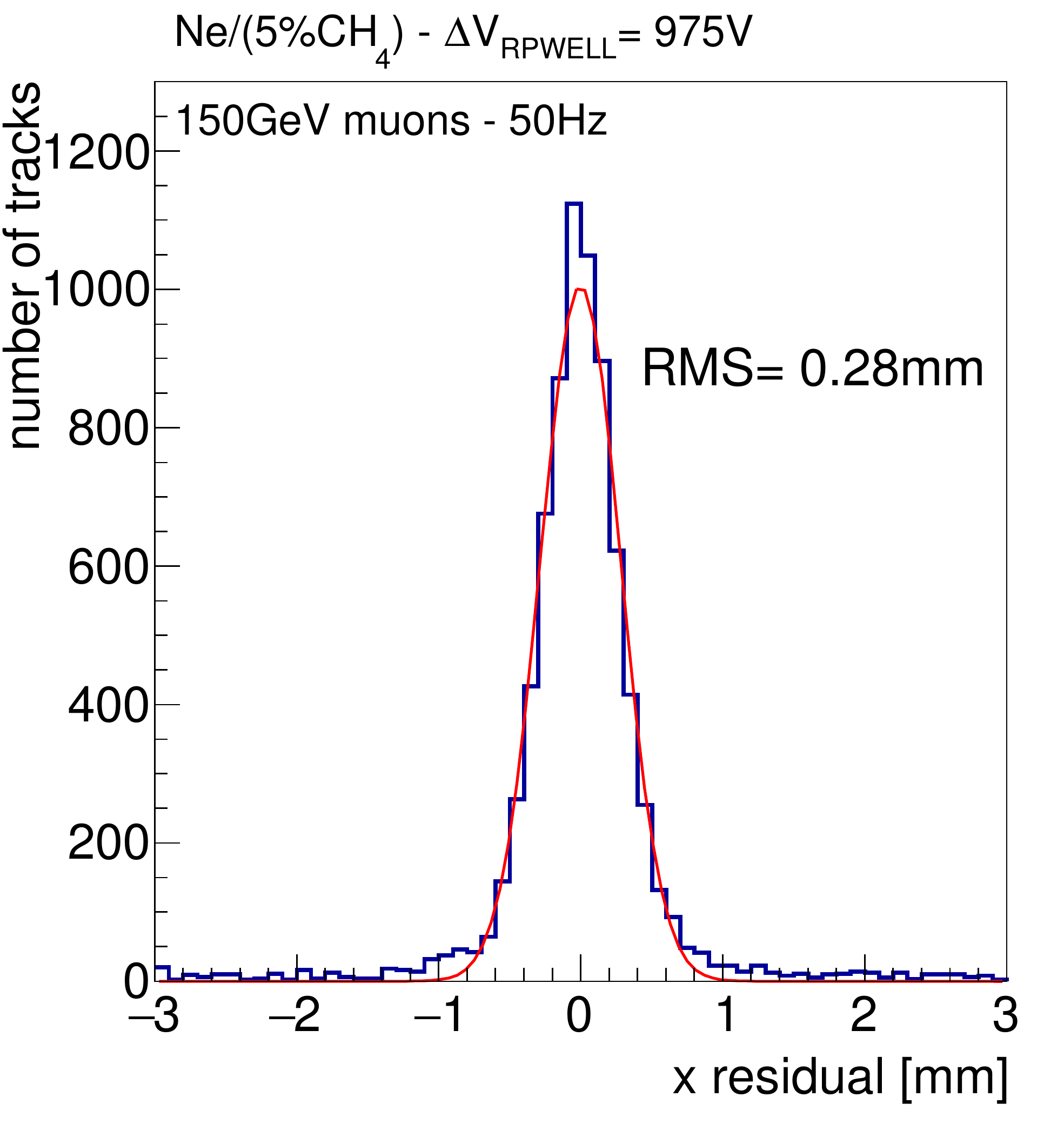}
\end{subfigure}
\begin{subfigure}[t]{0.5\textwidth}\caption{}
\includegraphics[scale=0.3]{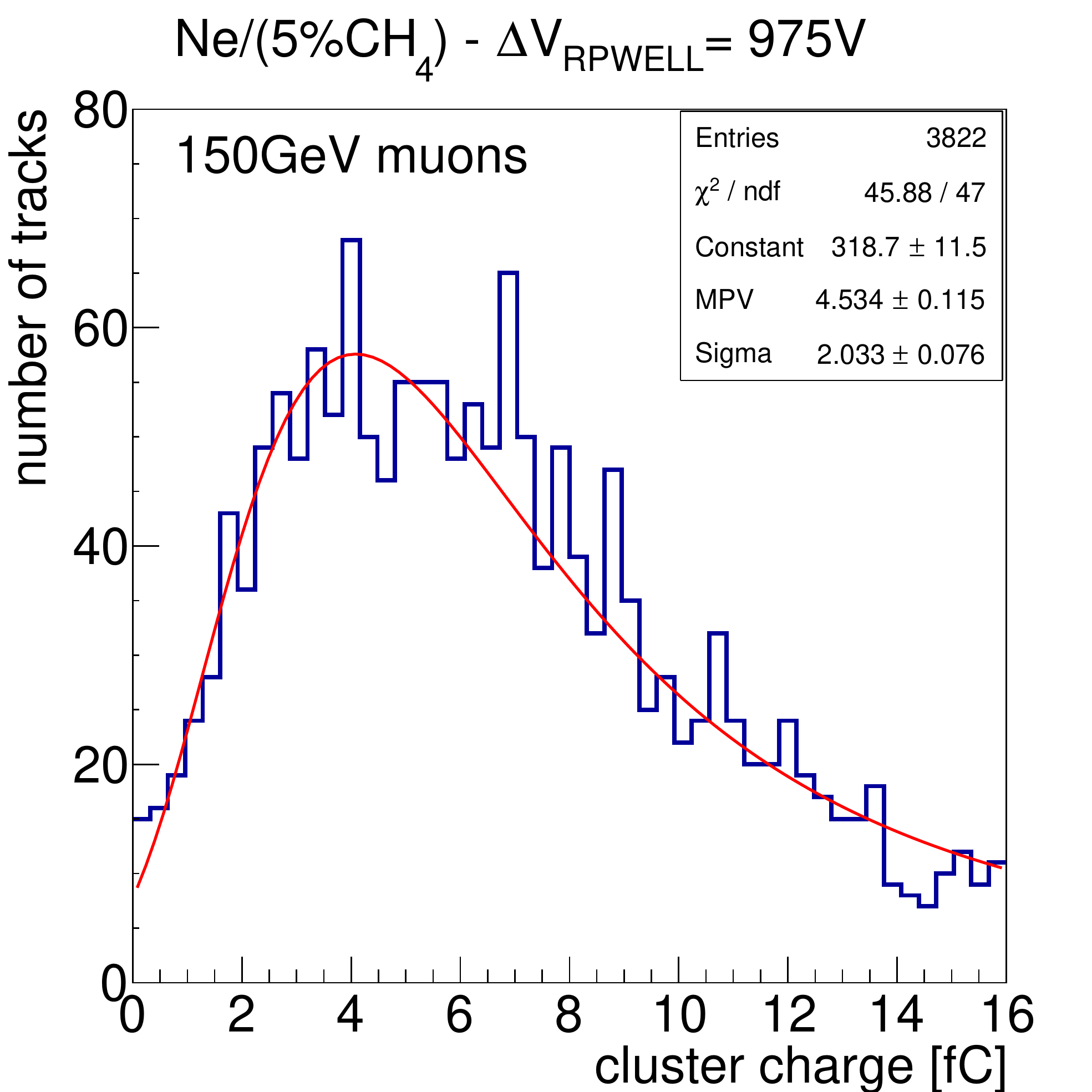}
\end{subfigure}
\caption{Residuals histogram (a) and calibrated cluster-charge spectrum (b) recorded with the detector shown in figure~\ref{fig: RPWELL detector}. The cluster-charge spectrum is fitted to a Landau function. The residuals histogram is fitted to a Gaussian, which RMS defines the detector position resolution (here 0.28mm RMS). \nech gas; \dvrpwell= 975~V (effective gain $\sim$3$\times$10$^3$); 1~mm strips pitch; 50~Hz 150~GeV/c muons at normal incidence. }\label{fig: spectrum - residuals histo}
\end{figure}

\begin{figure}[h]
\begin{subfigure}[t]{0.5\textwidth}\caption{}
\includegraphics[scale=0.3]{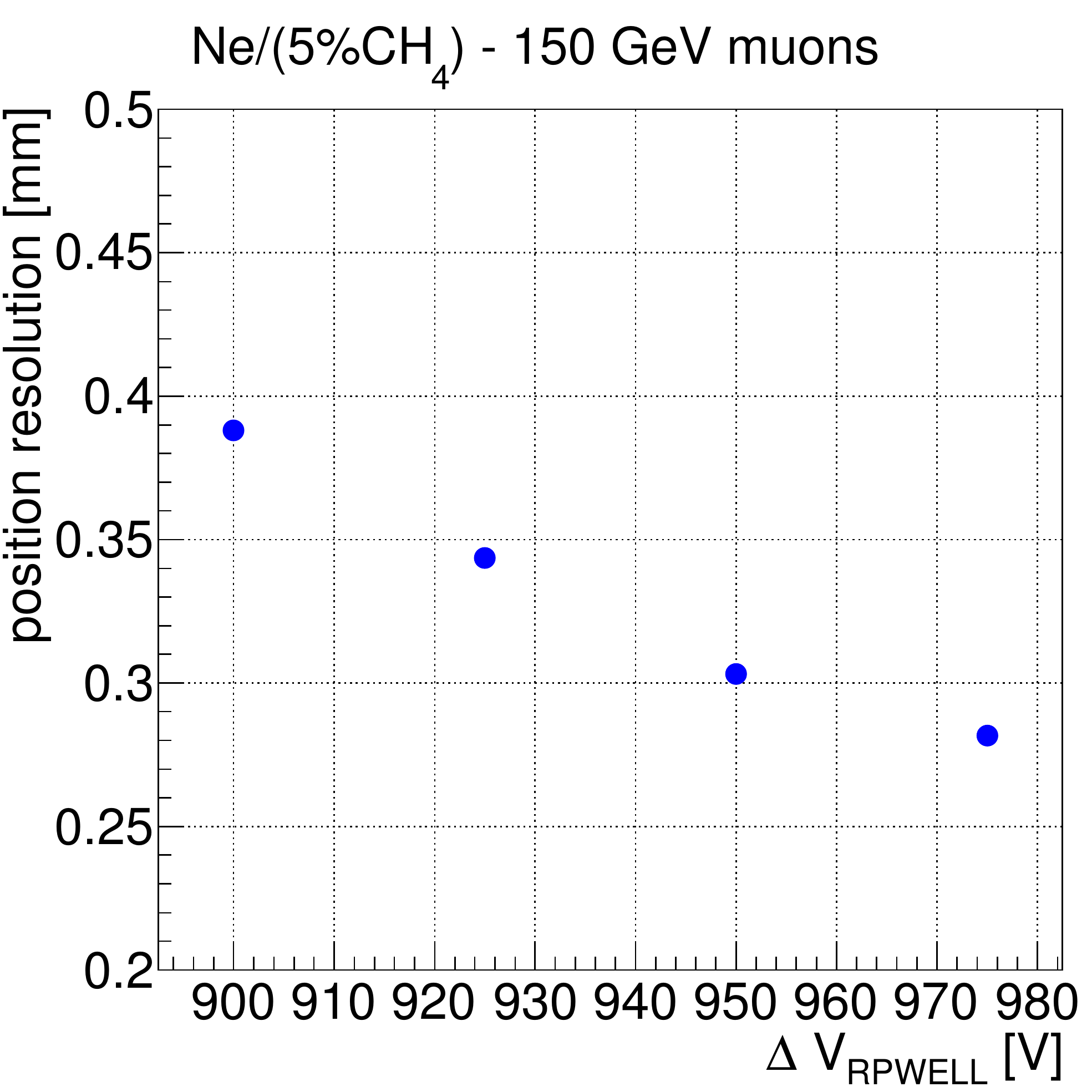}
\end{subfigure}
\begin{subfigure}[t]{0.5\textwidth}\caption{}
\includegraphics[scale=0.3]{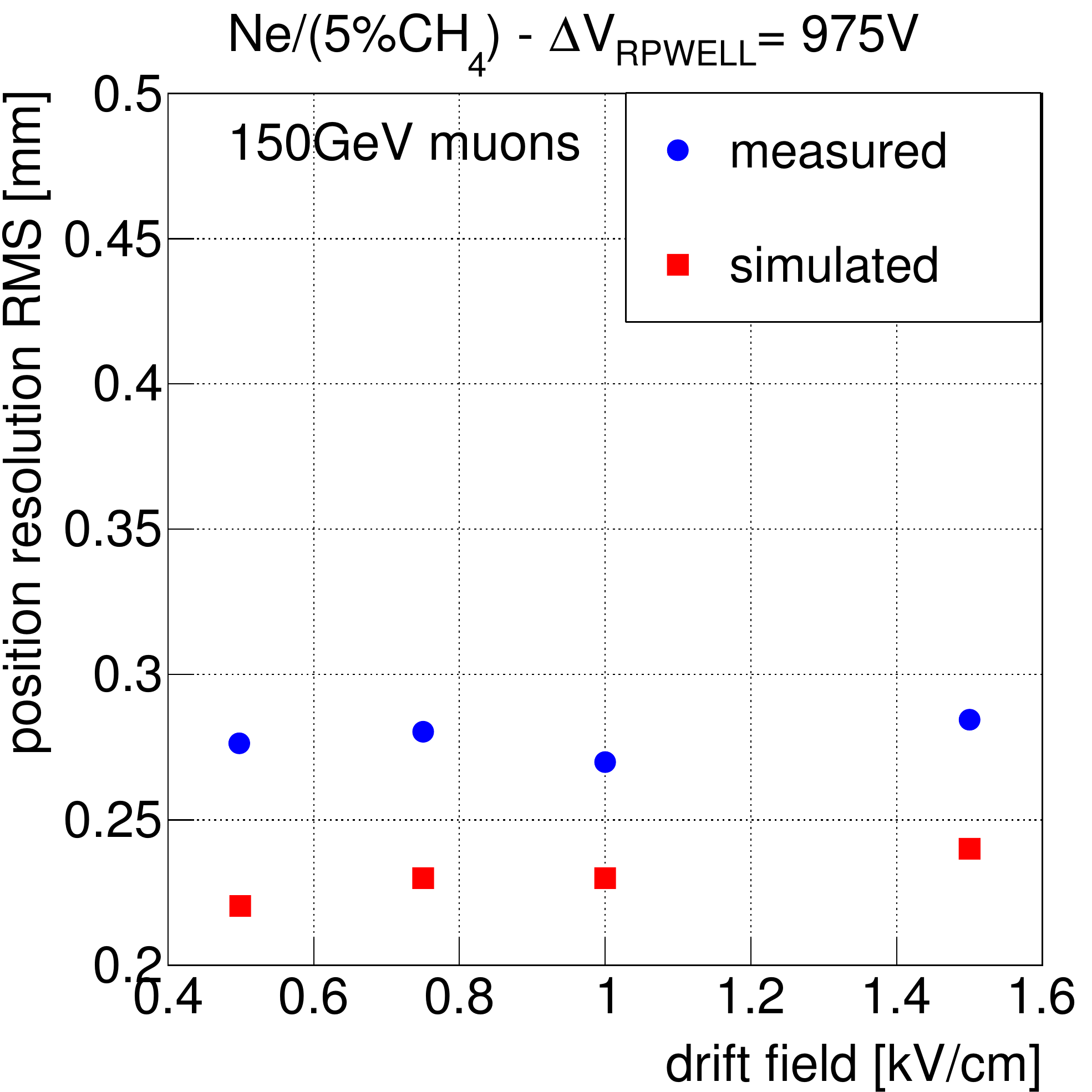}
\end{subfigure}
\caption{a) The measured RPWELL position resolution (residuals RMS) as a function of \dvrpwell.  b) The measured and simulated (see section~\ref{sec: Simulation}) position resolution at the maximum reachable value of \dvrpwell~ (effective gain $\sim$3$\times$10$^3$) for different values of the drift field. \nech gas; 1~mm strips pitch; 50~Hz 150~GeV/c muons at normal incidence. }\label{fig: HV scan - drift scan}
\end{figure}

The best position resolution of the RPWELL detector was obtained at the maximum achievable voltage \dvrpwell = 975~V\footnote{Note that as in~\cite{bressler2016first} (in similar conditions but with a different RP material), \dvrpwell could not be raised above 975~V, due to the appearance of small discharges, at the nA level.}, corresponding to an effective gain of $\sim$3$\times$10$^3$. In figure~\ref{fig: spectrum - residuals histo}-a we show the residuals histogram with its Gaussian fit; the corresponding RMS-value of the distribution is 0.28~mm. In figure~\ref{fig: spectrum - residuals histo}-b the cluster-charge spectrum for the same measurement is fitted to a Landau function.  At lower voltage values the localization properties of the detector degraded, as shown in figure~\ref{fig: HV scan - drift scan}-a were the residuals histogram RMS is plotted as a function of \dvrpwell. This can be explained as follows: position resolution better than the hole spacing is obtained only if the charge is shared between several holes. A better signal-to-noise ratio due to higher operation voltages gives a better sensitivity also to avalanches starting from small primary charge, resulting in effectively higher charge sharing between holes, and therefore improved position resolution. The relationship between position resolution and holes multiplicity is explained in more detail below and in section~\ref{sec: Simulation}. 

\begin{figure}[h]
\centering
\includegraphics[scale=0.7]{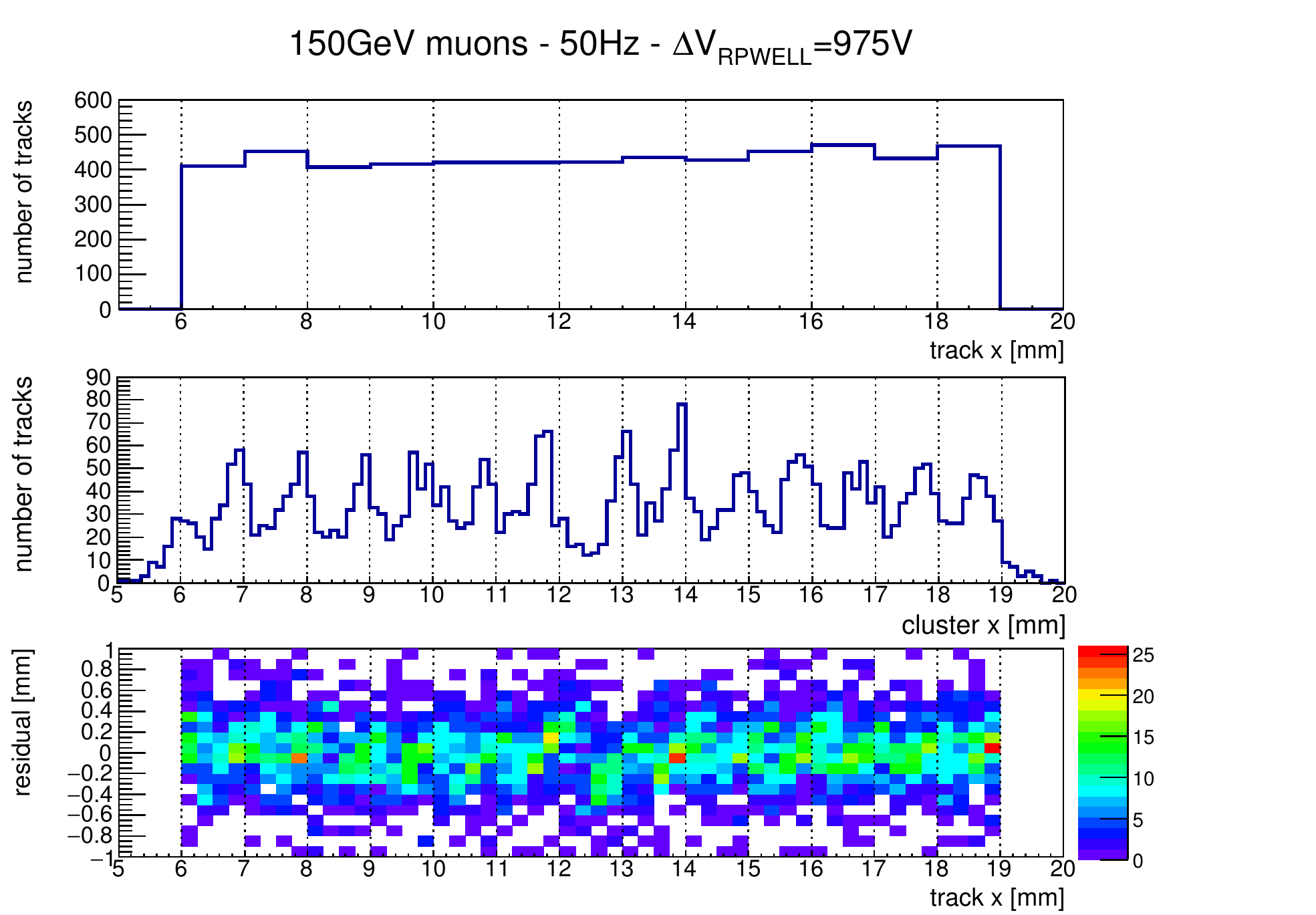}
\caption{The reconstructed muon-beam profile along the x-axis measured by the tracker (top) is compared to the reconstructed beam profile recorded by the RPWELL detector (middle). (bottom) Local-residuals pattern for the same data-set. The peaks in RPWELL-detector distribution correspond to the holes locations. \nech gas; \dvrpwell= 975~V(effective gain $\sim$3$\times$10$^3$); 1~mm strips pitch; 50~Hz 150~GeV/c muons at normal incidence.}\label{fig: profile - local residuals}
\end{figure}

\begin{figure}[h]
\begin{subfigure}[t]{0.5\textwidth}\caption{}
\includegraphics[scale=0.3]{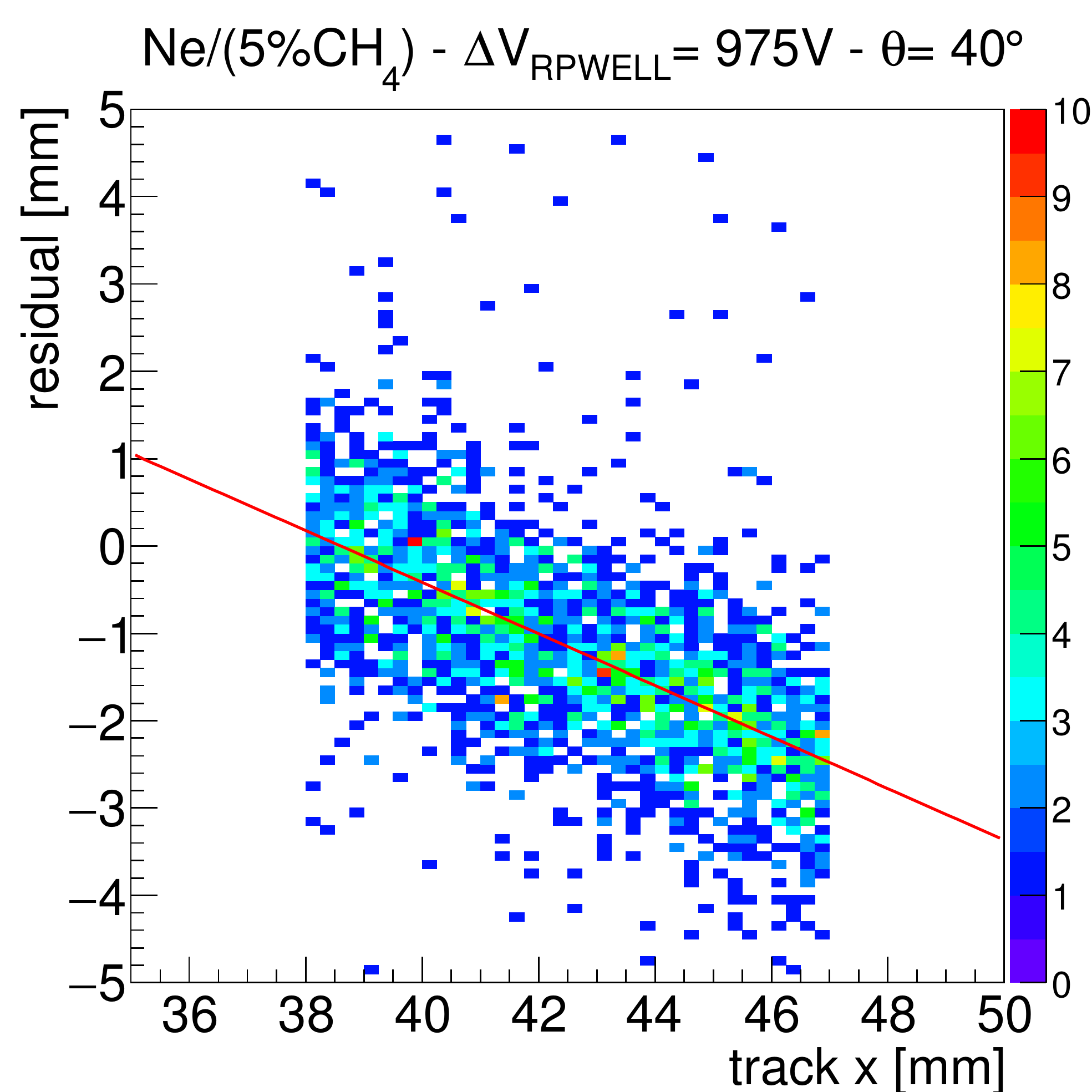}
\end{subfigure}
\begin{subfigure}[t]{0.5\textwidth}\caption{}
\includegraphics[scale=0.3]{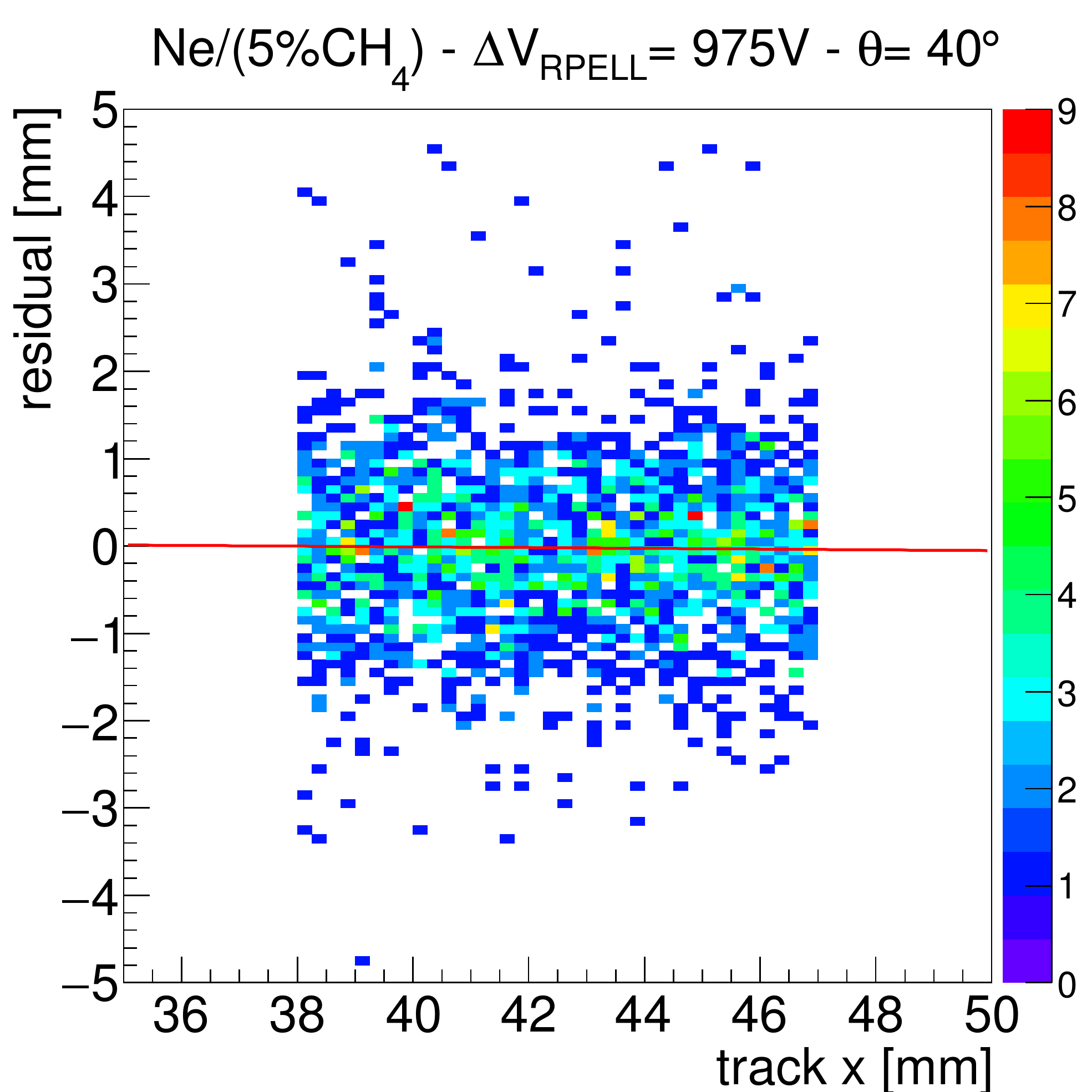}
\end{subfigure}
\caption{The local residual value vs particle track location at an incidence angle of $\uptheta$= 40$^\circ$, before (a) and after (b) linear correction. \nech gas; \dvrpwell= 975~V(effective gain $\sim$3$\times$10$^3$); 1.5~mm strips pitch; 50~Hz 150~GeV/c muons.}\label{fig: angle correction}
\end{figure}

\begin{figure}[h]
\begin{subfigure}[t]{0.5\textwidth}\caption{}
\includegraphics[scale=0.3]{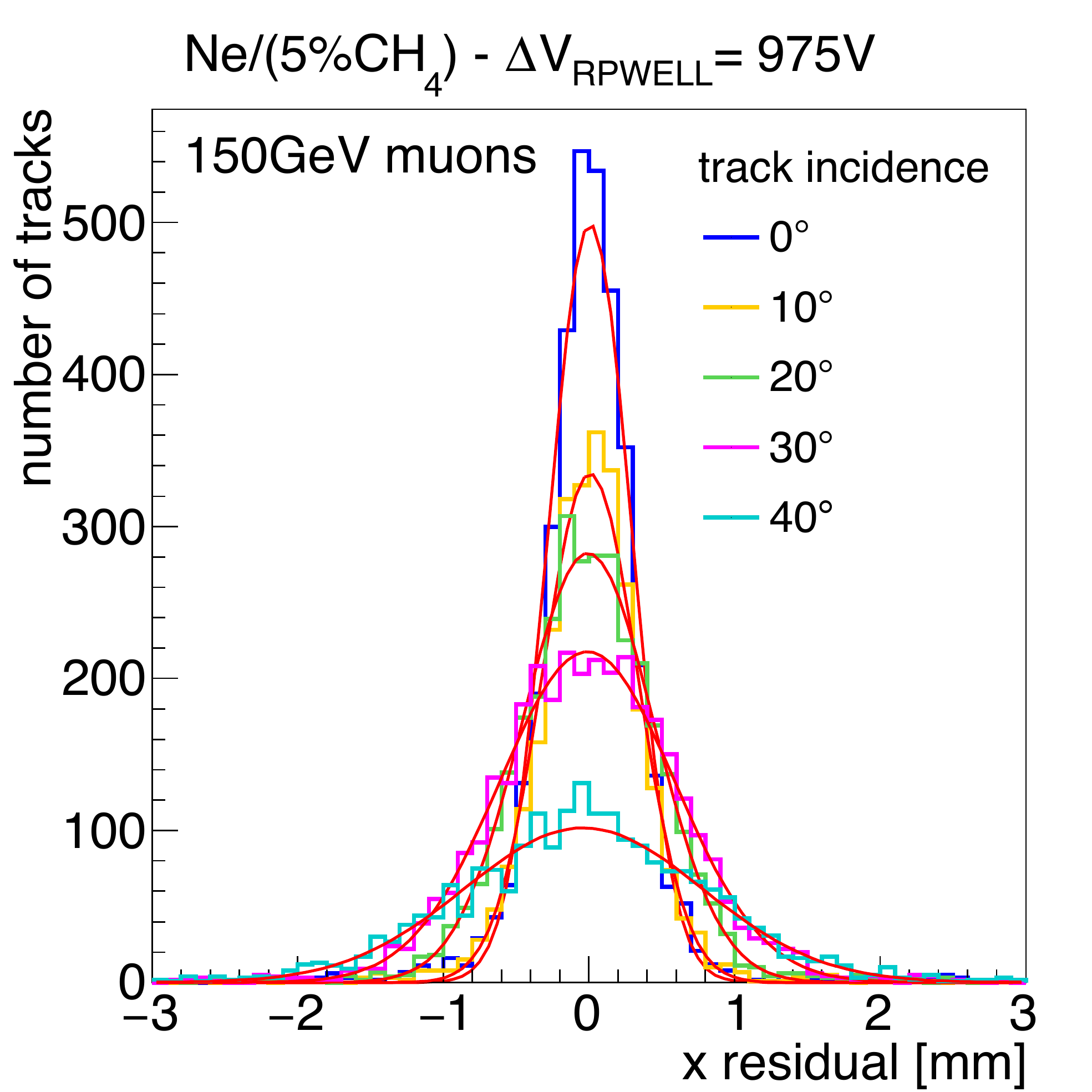}
\end{subfigure}
\begin{subfigure}[t]{0.5\textwidth}\caption{}
\includegraphics[scale=0.3]{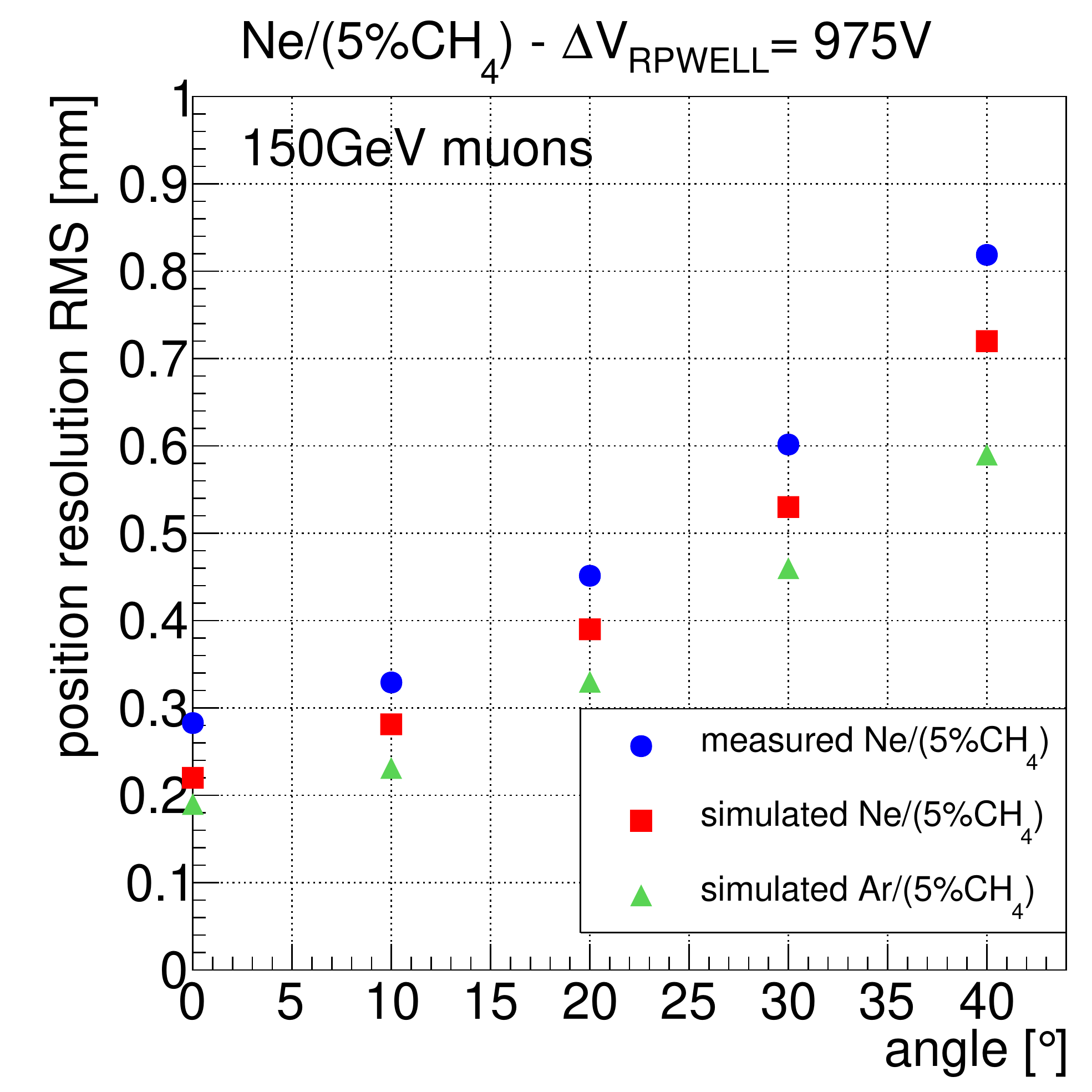}
\end{subfigure}
\caption{Distributions of the residuals (a); measured and simulated (see section~\ref{sec: Simulation}) RMS position resolution (b) for different particle-incident angles. \nech gas; \dvrpwell= 975~V (effective gain $\sim$3$\times$10$^3$); 1.5~mm strips pitch; 50~Hz 150~GeV/c muons.}\label{fig: angle scan}
\end{figure}

The effect of the drift field on the position resolution was found negligible as shown in figure~\ref{fig: HV scan - drift scan}-b together with the results of the simulation explained in section~\ref{sec: Simulation}. This suggests that the transverse electron diffusion in \nech does not contribute significantly to the detector performance.  
Looking at the local detector response suggests that the THGEM-electrode geometry plays a significant role in determining the detector's position resolution. Figure~\ref{fig: profile - local residuals}-top shows the reconstructed muon beam distribution along the x-axis as measured by the tracker for the same data set as in figure~\ref{fig: spectrum - residuals histo}. The equivalent measurement by the RPWELL detector is depicted in figure~\ref{fig: profile - local residuals}-middle. The measured RPWELL-detector distribution clearly reproduces the THGEM-holes pattern shown in figure~\ref{fig: RPWELL detector}-b. For the same measurement, figure~\ref{fig: profile - local residuals}-bottom shows a 2-D representation of the measured residuals as a function of the track position. The local residuals pattern can be explained as follows: for particles impinging at normal incidence (90$^\circ$ to the detector plane) in the position of a hole center, most of the ionization electrons along a track are focused into the same hole (except for occasional delta electrons and other electrons getting large transverse diffusion), where an avalanche develops inducing a signal. The reconstructed event position in this case is close to the  center of the hole, resulting in a residual value close to 0. The residual grows linearly with the track distance from the hole-center, due to an increasing number of electrons directed to a neighboring hole. At the limit, for tracks hitting the central region in-between two holes, the primary charge is shared - on average - equally among them, and the residual is again close to 0. In section~\ref{sec: Simulation} we study this effect in more detail with a dedicated model simulation.
We also studied the case of an angular particle incidence, in which the ionization electrons within the drift gap (here 5~mm) are, in most cases, collected by more than a single hole. The position resolution was evaluated at different track incidence angles, rotating the chamber around the y-axis from 0$^\circ$ to 40$^\circ$. The rotation had to be corrected for when determining the track position on the x-y plane: a particle traversing the detector perpendicularly to its plane ($\uptheta$= 0$^\circ$), at a point x, will traverse it at a point x' when the detector is rotated by an angle $\uptheta$, with x'= x/cos$\uptheta$. This appears clearly in the local-residuals plot, as shown in figure~\ref{fig: angle correction}-a for $\uptheta$= 40$^\circ$. To correct for this effect and correctly measure the position resolution we fitted the local-residuals plot to a line that was then subtracted from each residual. The result of this correction is shown in figure~\ref{fig: angle correction}-b. The typical local-residuals pattern shown in figure~\ref{fig: profile - local residuals}-bottom for $\uptheta$= 0$^\circ$ vanishes at large angles, since the fraction of primary charges reaching each hole is not anymore correlated uniquely with the track position. This effect results also in a degradation of the position resolution, as reflected in the residuals histograms plotted in figure~\ref{fig: angle scan}-a for different incidence angles. In figure~\ref{fig: angle scan}-b we show the measured position resolution as a function of the incidence angle together with the results of the simulation explained in section~\ref{sec: Simulation}. At 40$^\circ$ the position resolution is 0.82~mm RMS - about a 4-fold degradation compared to the perpendicular incidence.

\section{Monte Carlo simulations}
\label{sec: Simulation}

The RPWELL detector operation was modeled by a Monte Carlo (MC) simulation. A simplified THGEM-electrode geometry was chosen, with the same parameters as indicated in section~\ref{sec: RPWELL}, but with a "homogeneous" array of holes - without the central 1.3~mm spacing (figure~\ref{fig: RPWELL detector}). Considering the number of holes involved in the measurement, this difference in geometry should not affect the results significantly. The simulation was performed on an event-by-event basis, and it included the physics processes detailed below. The detector performance was reproduced for a set of 20,000 events generated randomly across the detector plane. To emulate the strip readout, all the estimated quantities were integrated along the y-axis and presented, when needed, as a function of the x-axis only.

\subsection{Event generator}
Using the Garfield simulation framework~\cite{veenhof2015garfield}, together with the neBEM solver~\cite{muhkopadhyay2006computation}, Heed~\cite{smirnov2005modeling} and Magboltz software~\cite{biagi2016magboltz}, we implemented a single-event generator. To make this method CPU-effective, we didn't simulate all the physics processes from first principles, but instead we used measured data as inputs whenever this was possible, in particular:
\begin{itemize}
\item Single-electron gain spectrum, measured with UV-photons in similar experimental conditions.
\item Cluster-charge spectrum, from the test-beam with muons (figure~\ref{fig: spectrum - residuals histo}-b).
\item Typical signal shapes from the readout strips, digitized by the SRS/APV25 electronics (figure~\ref{fig: induced signal}).
\end{itemize}

For each event we considered the following physics processes:
\begin{enumerate}
\item \underline{Primary-electron clusters generation} by a 150~GeV muon traversing the detector, distributed along a track within the drift gap. This process was simulated as described in appendix~\ref{sec: PE library}.
\item \underline{Primary-electron drift and diffusion} under electric field towards the THGEM and their focusing into the holes. 
\item \underline{Primary-electron multiplication} in the high-field region of the multiplier holes by avalanche process. 
\item \underline{Current-signal induction on the anode strips} by the drift of avalanche-generated electrons and ions towards the anode and the THGEM top respectively. 
The long ($\sim$2~$\upmu$s) current pulse was shaped and digitized by the readout electronics, with a shaping time of $\sim$100~ns. While in these simulations we used as input the measured signal (see details in appendix~\ref{sec: appendix}), investigating the signal formation process in detail will be part of another dedicated study.
\item \underline{Event position reconstruction} by the charge-weighted signals. Residuals are defined as the difference between the original event position and the reconstructed one.
\end{enumerate}

More details on the single-event generator are given in the appendix. 

\subsection{Simulation results}
\label{sec: simulation results}

The simulations were performed for the detector's operation conditions yielding the best experimental position resolution, namely with \dvrpwell= 975~V.

\begin{figure}[h]
\centering
\includegraphics[scale=0.3]{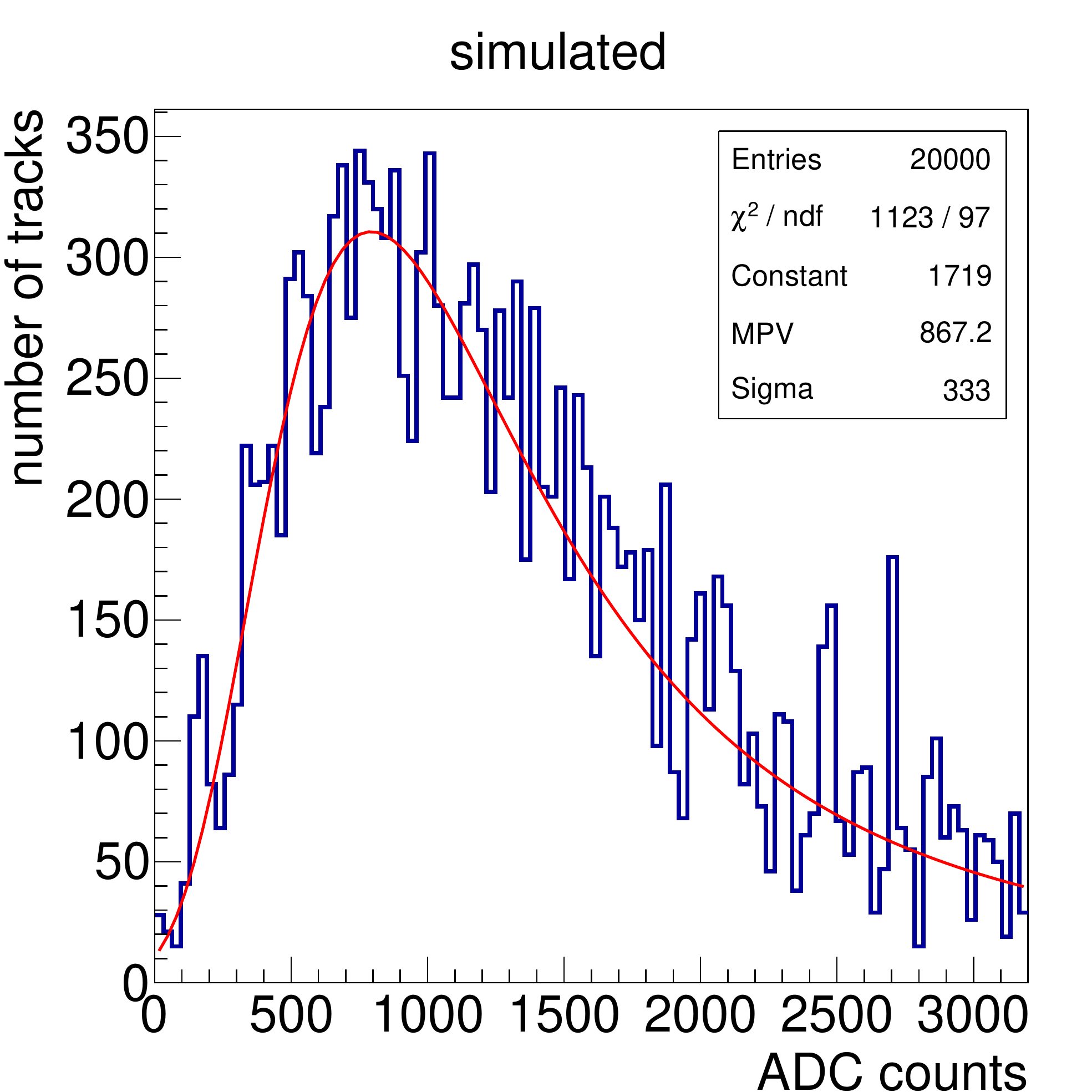}
\caption{Simulated RPWELL cluster charge-spectra for muons at normal incidence fitted to a Landau function with the indicated most probable value (MPV) and sigma parameter. \nech gas; effective gain $\sim$3$\times$10$^3$; 1~mm strips pitch.}\label{fig: simulated spectrum}
\end{figure}

\begin{figure}[h]
\begin{subfigure}[t]{0.45\textwidth}\caption{}
\includegraphics[scale=0.3]{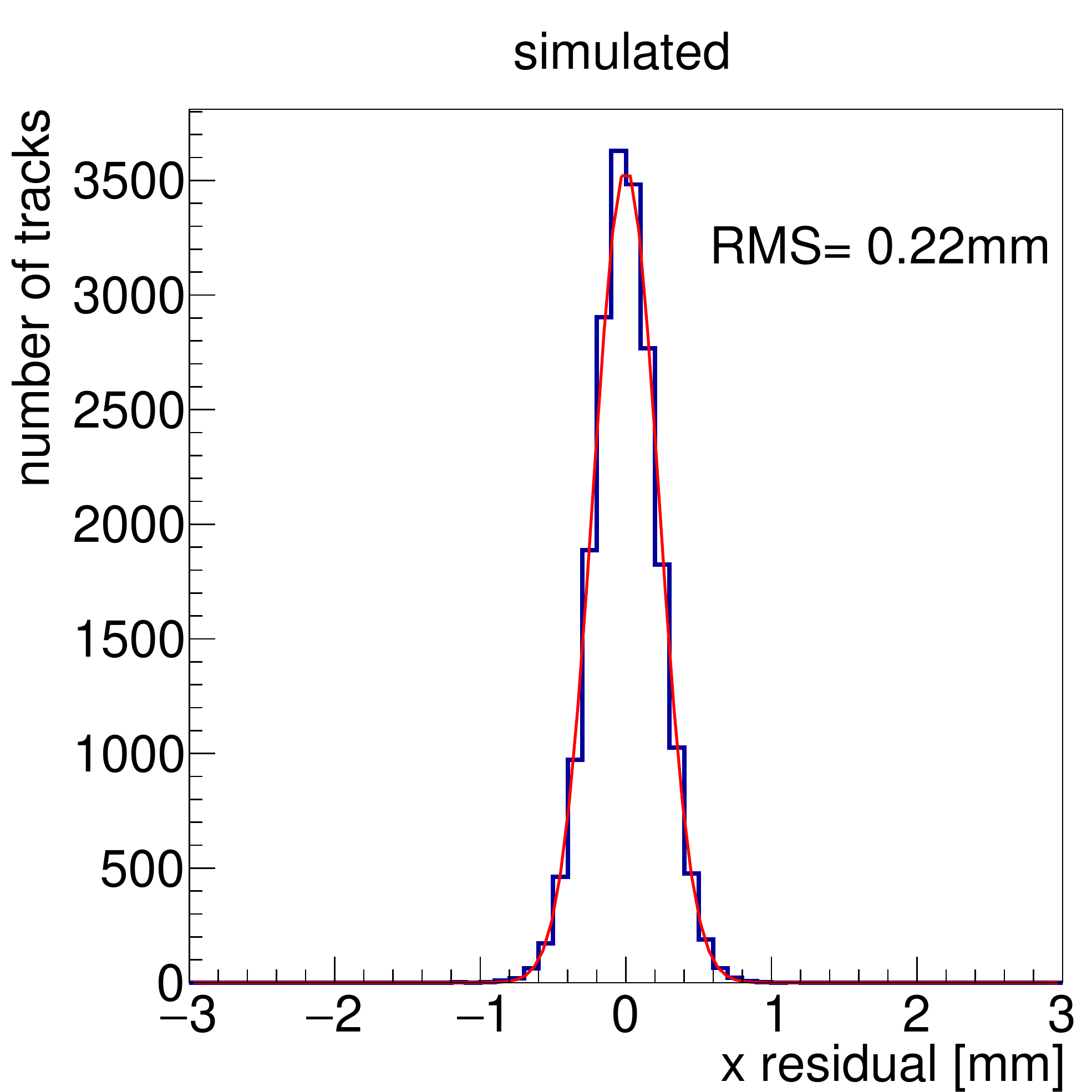}
\end{subfigure}
\begin{subfigure}[t]{0.6\textwidth}\caption{}
\includegraphics[scale=0.45]{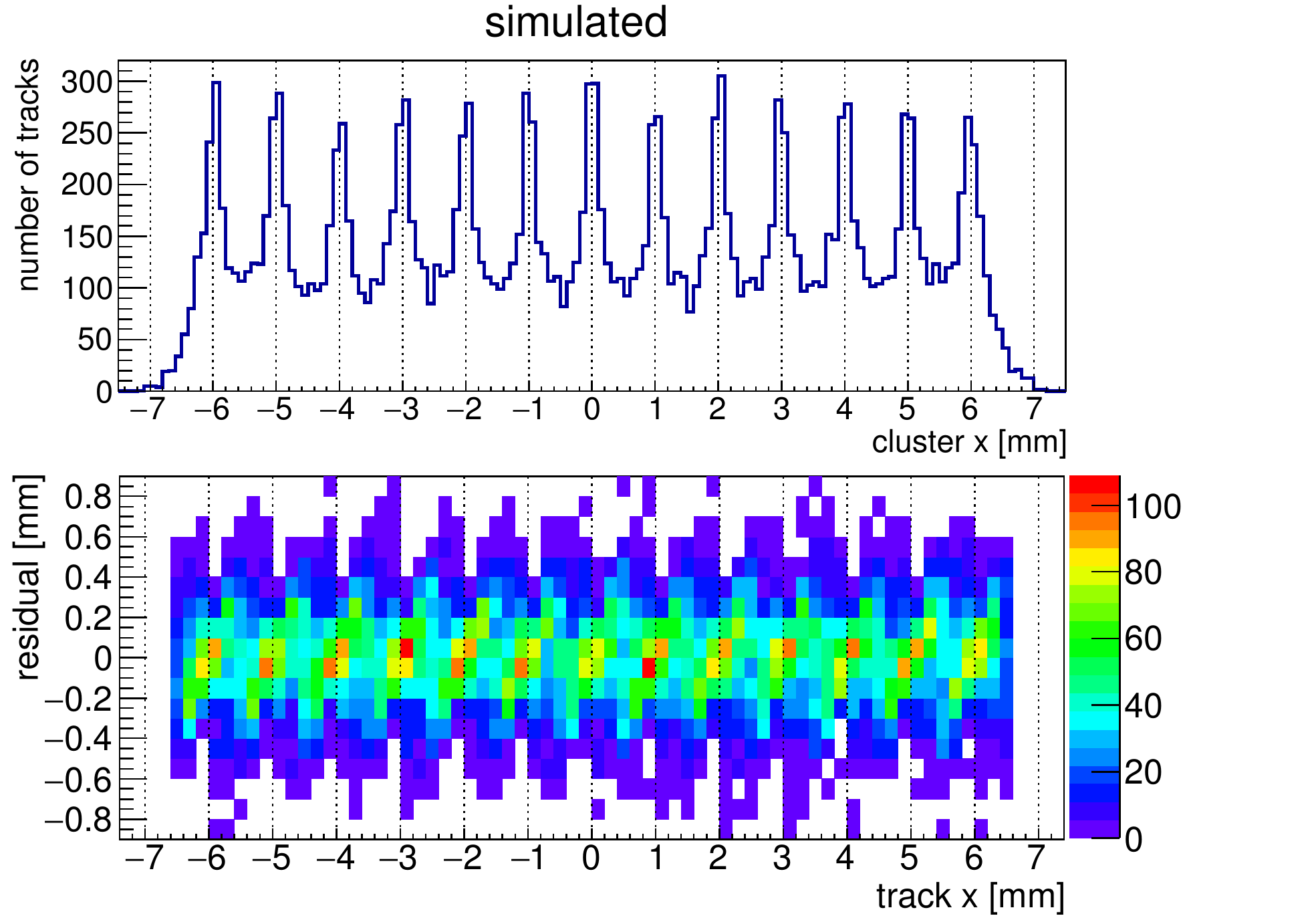}
\end{subfigure}
\caption{Simulation of the RPWELL operated in \nech mixture detecting 150~GeV muons at normal incidence: a) Residuals histogram, resulting in 0.22~mm RMS position resolution value. b) Simulated detector response. Reconstructed clusters position histogram along the x-axis (top), and local residuals pattern (bottom). Effective gain $\sim$3$\times$10$^3$; 1~mm strips pitch.}\label{fig: simulation residuals - profile}
\end{figure}

\begin{figure}[h]
\begin{subfigure}[t]{0.5\textwidth}\caption{}
\includegraphics[scale=0.3]{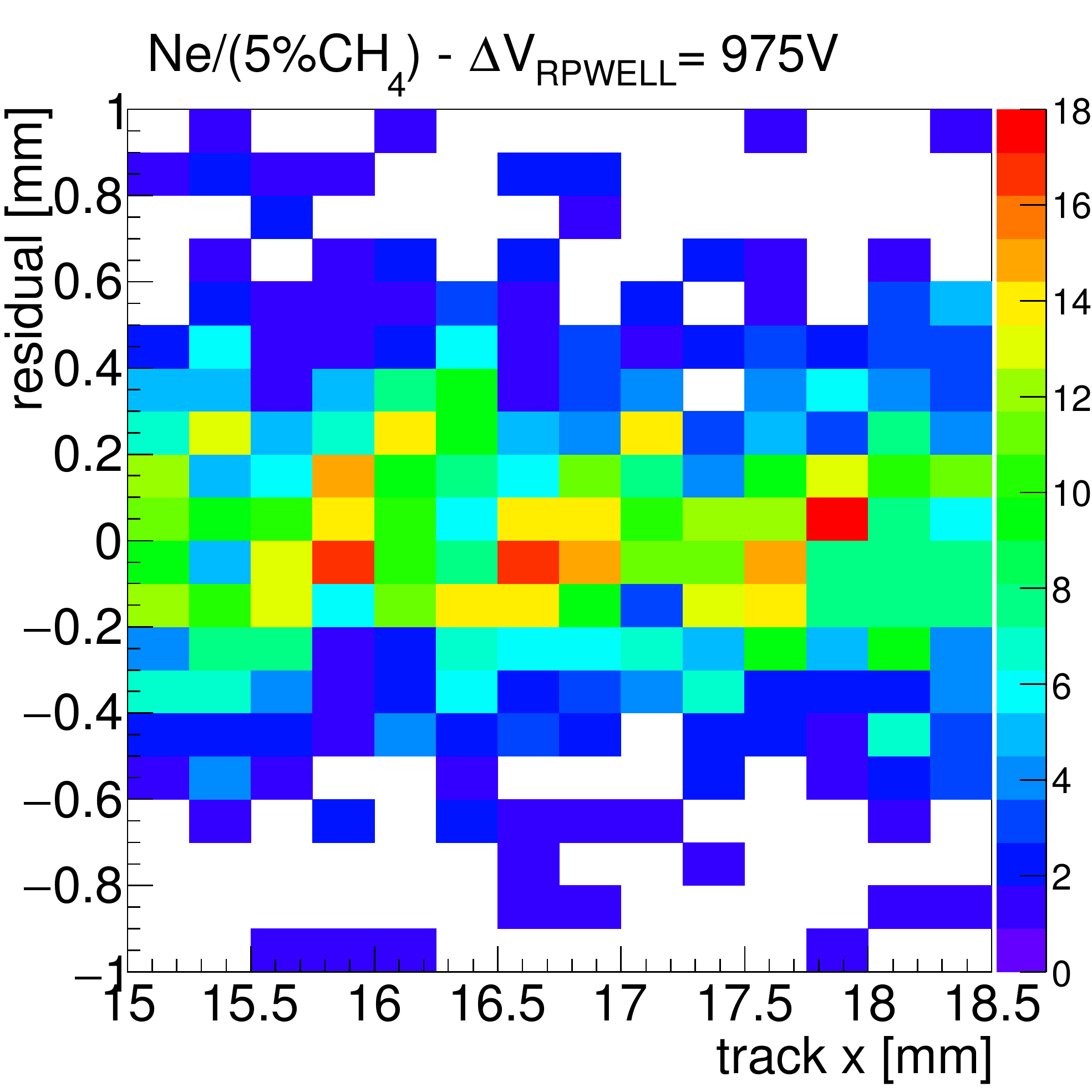}
\end{subfigure}
\begin{subfigure}[t]{0.5\textwidth}\caption{}
\includegraphics[scale=0.3]{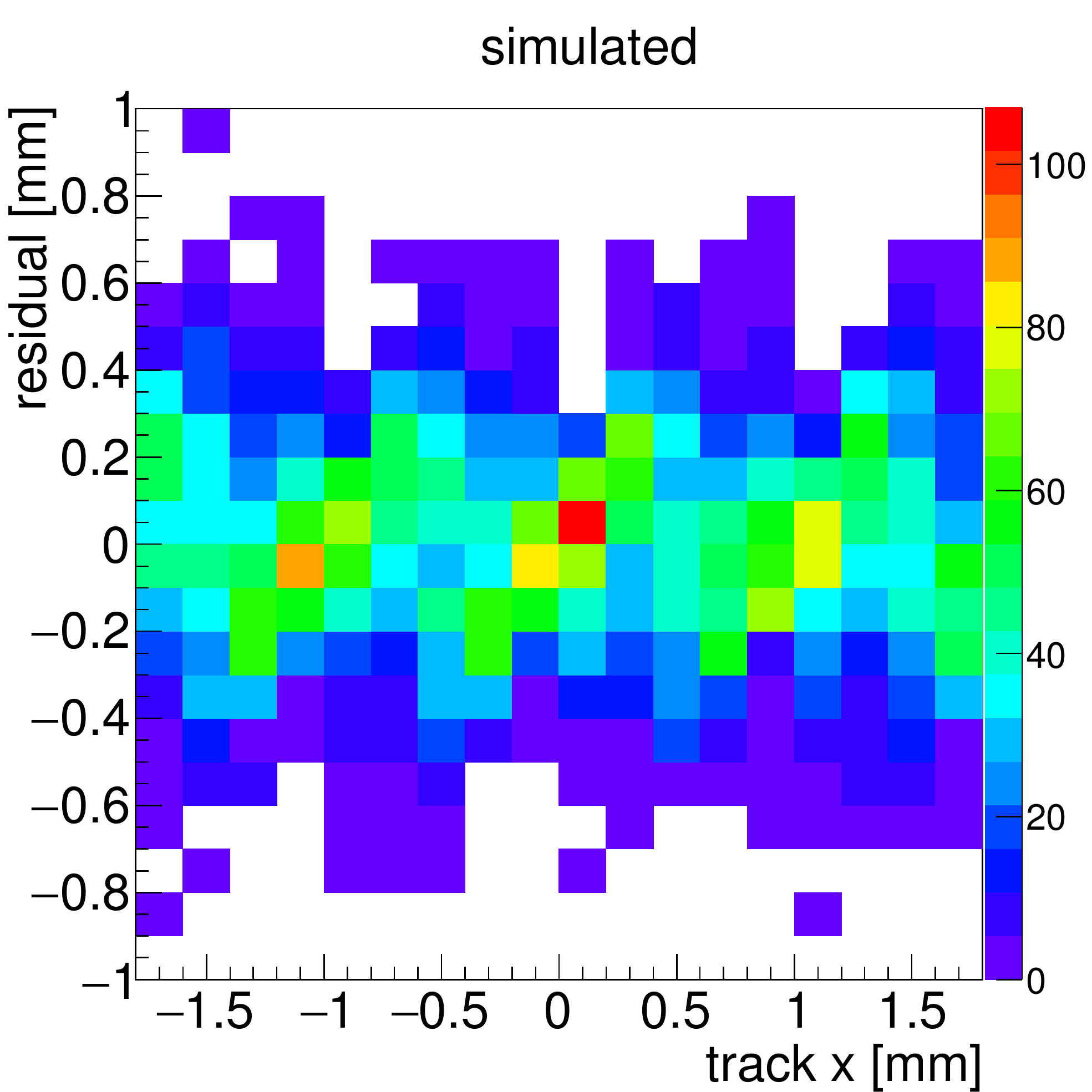}
\end{subfigure}
\caption{Local residual patterns for muons at normal incidence in the RPWELL operated in \nech mixture: measured (a) and simulated (b). Effective gain $\sim$3$\times$10$^3$; 1~mm strips pitch.}\label{fig: local residuals zoom}
\end{figure}

Figure~\ref{fig: simulated spectrum} shows the Landau fit of the simulated charge spectrum. It is in good agreement with the measured one (figure~\ref{fig: spectrum - residuals histo}-b). Figure~\ref{fig: simulation residuals - profile}-a shows the simulated residuals histogram; the position resolution of 0.22~mm RMS is superior to the experimental one of 0.28~mm shown in figure~\ref{fig: spectrum - residuals histo}-a. The small difference could be due to detector noise and gain non-uniformity.
The simulated profile of the reconstructed track-position (top panel of figure~\ref{fig: simulation residuals - profile}-b) is in good agreement with the experimental distribution shown in the middle panel of figure~\ref{fig: profile - local residuals}. In both cases, the peaks are correlated with the THGEM holes, due to charge focusing into holes. The characteristic pattern of the residuals as a function of the event position, due to primary-charge focusing into the holes (bottom panel of figure~\ref{fig: profile - local residuals}), is clearly visible in the simulated data (bottom panel of figure~\ref{fig: simulation residuals - profile}-b). In figure~\ref{fig: local residuals zoom} we compare the simulated local residuals with the measured ones in a small detector region. The patterns are very similar, with the residuals reaching zero values in correspondence to THGEM holes and the central region between holes; the latter due to charge sharing (see figure~\ref{fig: charge sharing} in the appendix).

\begin{figure}[h]
\begin{subfigure}[t]{0.5\textwidth}\caption{}
\includegraphics[scale=0.3]{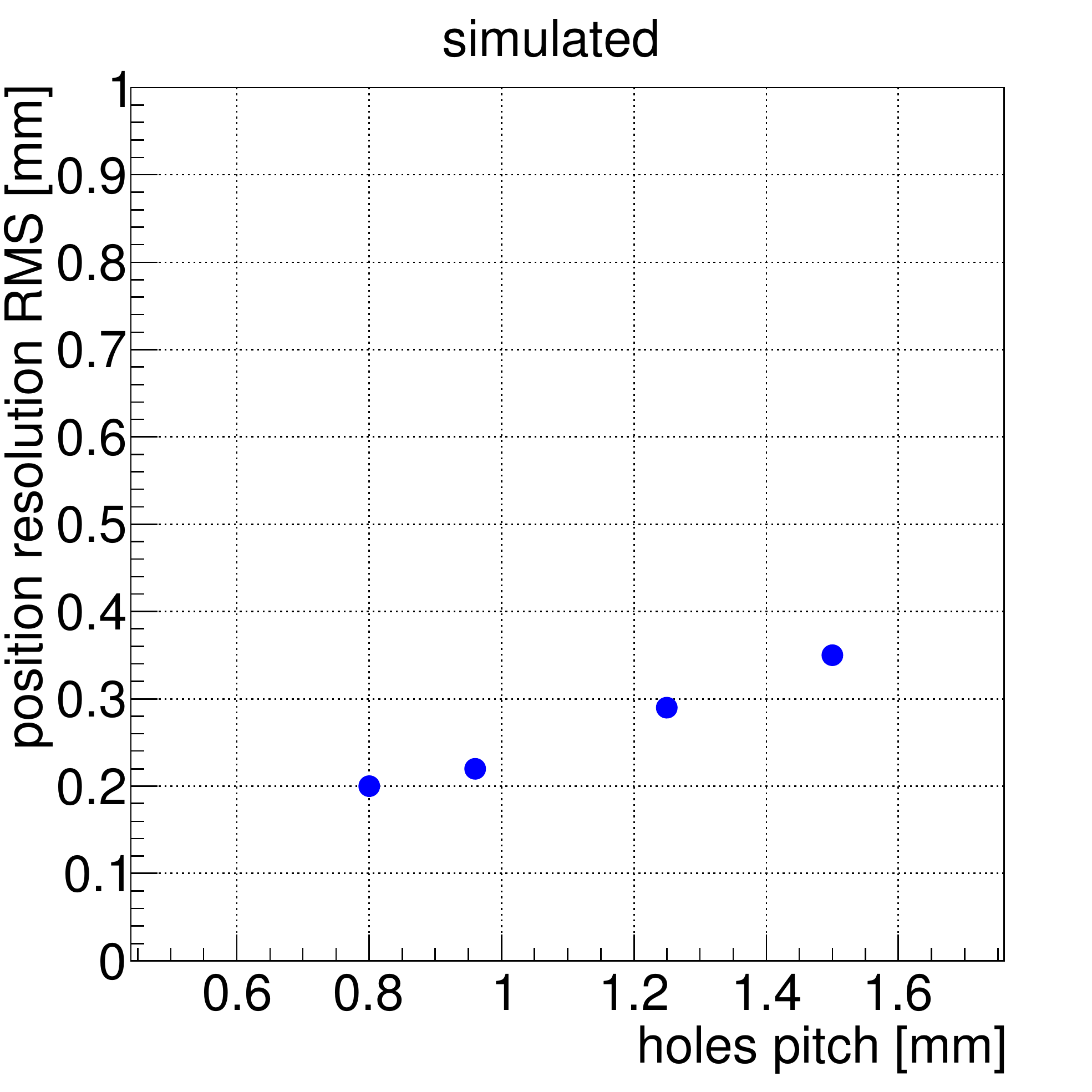}
\end{subfigure}
\begin{subfigure}[t]{0.5\textwidth}\caption{}
\includegraphics[scale=0.3]{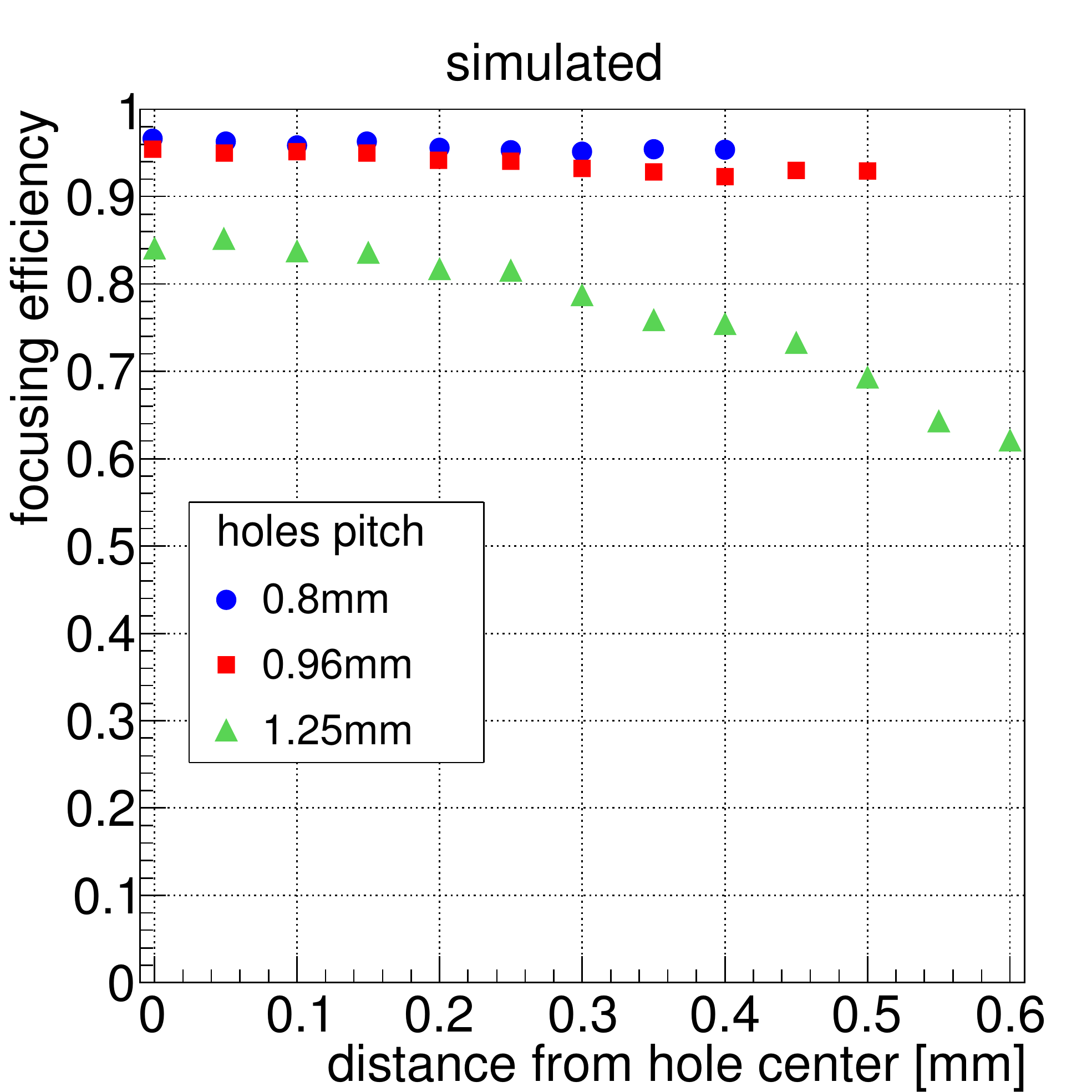}
\end{subfigure}
\caption{Simulated RPWELL position resolution for muons at normal incidence (a) and electron focusing efficiency into the THGEM holes (b) as a function of the THGEM holes pitch in \nech mixture. Effective gain $\sim$3$\times$10$^3$; 1~mm strips pitch.}\label{fig: simulated pitch scan}
\end{figure}

We could also reproduce the position resolution dependence on the incidence angle and on the drift field, as can be seen in figure~\ref{fig: HV scan - drift scan}-b and \ref{fig: angle scan}-b. The simulation confirms the measured degradation of the position resolution at increasing angles and the negligible effect of the drift field up to 1.5~kV/cm.
To demonstrate that the detector localization properties strongly depend on the THGEM hole pitch, we simulated the position resolution for different pitch values. Varying the pitch from 0.8~mm to 1.5~mm resulted in an almost two-fold degradation in the position resolution (figure~\ref{fig: simulated pitch scan}-a). In the same pitch range, the detector gain was not affected, but we could see some primary electron losses at the largest hole-pitch value (figure~\ref{fig: simulated pitch scan}-b).
Aiming at improving the RPWELL localization properties, we simulated the performance of the detector operating in \arch. For this simulation we kept the same parameters as for \nech, except for the primary-electron library (see the appendix), that was recalculated for \arch with \dvrpwell= 1800~V, giving an effective gain of $\sim$3$\times$10$^3$ ( similar to the one in \nech at \dvrpwell= 975~V).
Because of the larger number of primary electrons produced by a MIP in argon with respect to neon, the fluctuations of the charge sharing among THGEM holes in this configuration were smaller, resulting in an improved position resolution - compared to \nech - of 0.19~mm RMS at normal incidence. The calculated RMS at incidence angles up to 40$^\circ$ is shown in figure~\ref{fig: angle scan}-b. Compared to neon, the degradation at increasing incidence angles is better, 3-times instead of 4-times. This is because in argon the primary electron distribution along the track is more uniform, preserving a better correlation between the particle position and the reconstructed cluster.

Good agreement is found between the measured detector performances and the simulated one. This indicates that the most relevant physics processes contributing to the position resolution were taken into account correctly.  

\section{Summary and discussion}
\label{sec: Summary and discussion}

Robust THGEM-based particle detectors were conceived for applications requiring particle tracking over large area, at moderate sub-millimeter localization resolutions. The localization properties of an RPWELL detector were investigated in this work for the first time. The detector comprises of a single-stage THGEM electrode, with 0.5~mm diameter holes, with 1~mm pitch; it is coupled to a readout plane with 1-D strips, via a glass resistive plate. Measurements with 150~GeV muons showed that the position resolution improves with increasing detector operation voltage (effective gain); this is attributed to the better signal-to-noise ratio causing an enhanced effective charge sharing among multiplier holes. The drift field in the conversion gap did not affect the resolution, indicating that in this configuration in \nech the electron transverse diffusion doesn't play a major role. The best position resolution measured in the present configuration, at normal incidence, is 0.28~mm RMS, at an effective gain $\sim$3$\times$10$^3$; it is about 4-fold smaller than the holes pitch. 
The position resolution deteriorated under an angular particle incidence. The measured resolution degraded from 0.28 to 0.8~mm RMS for respective incidence angles of 0$^\circ$ to 40$^\circ$. The main factor determining this degradation is that the primary electrons are produced in several clusters randomly distributed along a track. At non-0 incidence angles, the fraction of primary charges reaching each hole is not correlated uniquely with the track position. A possible way to contrast this effect would be using higher gas densities, obtaining a more uniform primary electron distribution along a track, and having a smaller drift gap, resulting in a smaller lateral primary charge distribution.

Monte Carlo simulations, incorporating the physics phenomena contributing to the position resolution, were carried out for predicting the expected detector performance. The simulation results reproduced rather well the experimental ones and confirm the prediction that the RPWELL electrode geometry plays a dominant factor in determining the position resolution.
This is very different from GEM-like structures where the hole size and pitch are significantly smaller, with primary charges being always distributed among several holes; this permits calculating their center-of-gravity - yielding position resolutions of a few tens of microns~\cite{lener2016mu,carnegie2005resolution}. 
The position resolution of the RPWELL can be improved if the charge sharing between the holes is improved. This can be achieved using different gaseous mixtures or optimizing the THGEM geometry.

\acknowledgments
We thank Professor Yi Wang (Tsinghua University, China) for providing us with the low-resistivity silicate glass samples. This research was supported in part by the I-CORE Program of the Planning and Budgeting Committee, the Nella and Leon Benoziyo Center for High Energy Physics, the Mel and Joyce Eisenberg-Keefer Fund for New Scientists and by Grant No 712482 from the Israeli Science Foundation (ISF).

\bibliographystyle{elsarticle-num}                 
\bibliography{bibliography.bib}

\appendix

\section{Monte Carlo single event generator}
\label{sec: appendix}
\begin{figure}[h]
\centering
\includegraphics[scale=0.3]{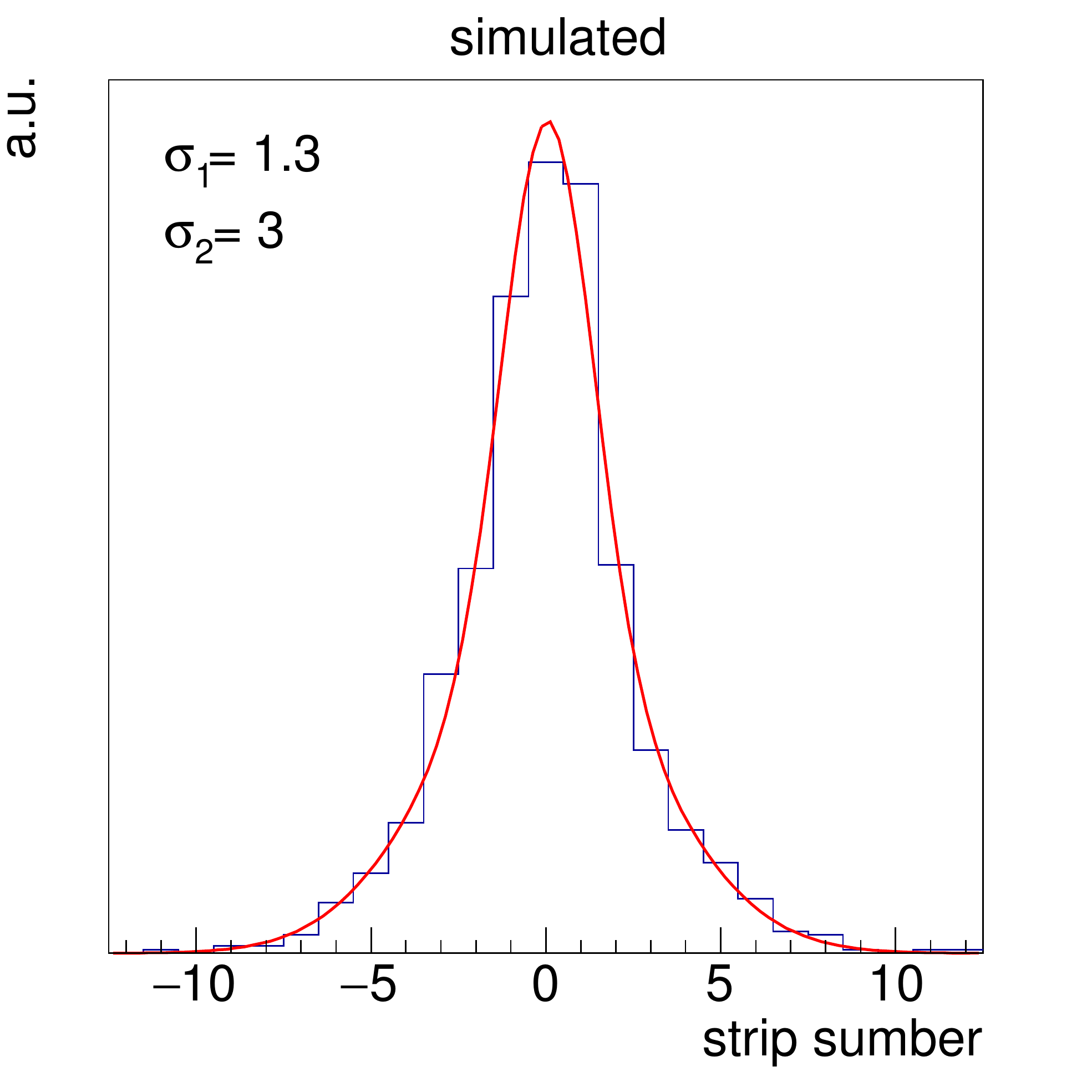}
\caption{Simulated digitized signal from the SRS/APV25 readout from strips.}\label{fig: simulated signal}
\end{figure}

The Monte Carlo simulation single-event generator is described. To make it CPU- effective, we didn't simulate all physics processes involved from first principles, but we rather used measured data as inputs whenever this was possible - in particular the following ones:
\begin{itemize}
\item Primary-electrons library (see section~\ref{sec: PE library} below).
\item Single-electron gain spectrum measured with UV photons in similar experimental conditions. This was fitted to an exponential.
\item Cluster-charge spectrum from the test-beam muons (figure~\ref{fig: spectrum - residuals histo}-b).
\item Typical signal shapes from the readout strips digitized by the SRS/APV25 electronics (figure~\ref{fig: induced signal}). We fitted it to a double Gaussian to consider the tails due to the charge spread on the resistive-plate surface. This effect is explained in~\cite{lin2014signal} for a resistive Micromegas detector and will be investigated in a future work. In figure~\ref{fig: simulated signal} we show an example of a simulated signal. 
\end{itemize}

We modeled the full detector geometry using the Garfield simulation framework~\cite{veenhof2015garfield}, together with the neBEM solver~\cite{muhkopadhyay2006computation}, Heed~\cite{smirnov2005modeling} and Magboltz software~\cite{biagi2016magboltz}. 
Below we describe the generation of the primary-electrons library and the single-event generator which contains all the steps of the Monte Carlo simulation.

\subsection{Primary electrons library}
\label{sec: PE library}

\begin{figure}[h]
\begin{subfigure}[t]{0.3\textwidth}\caption{}
\includegraphics[scale=0.2]{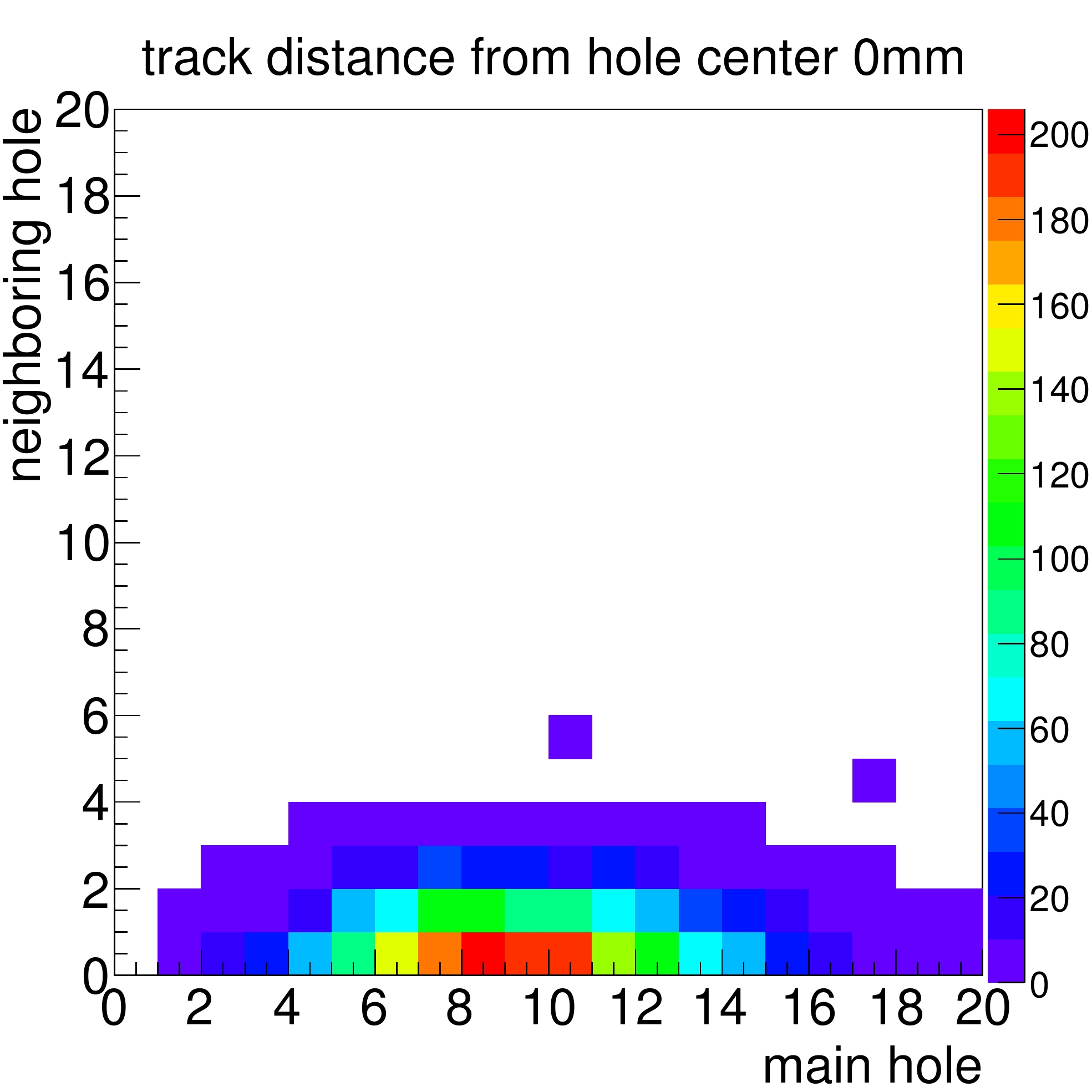}
\end{subfigure}
\begin{subfigure}[t]{0.3\textwidth}\caption{}
\includegraphics[scale=0.2]{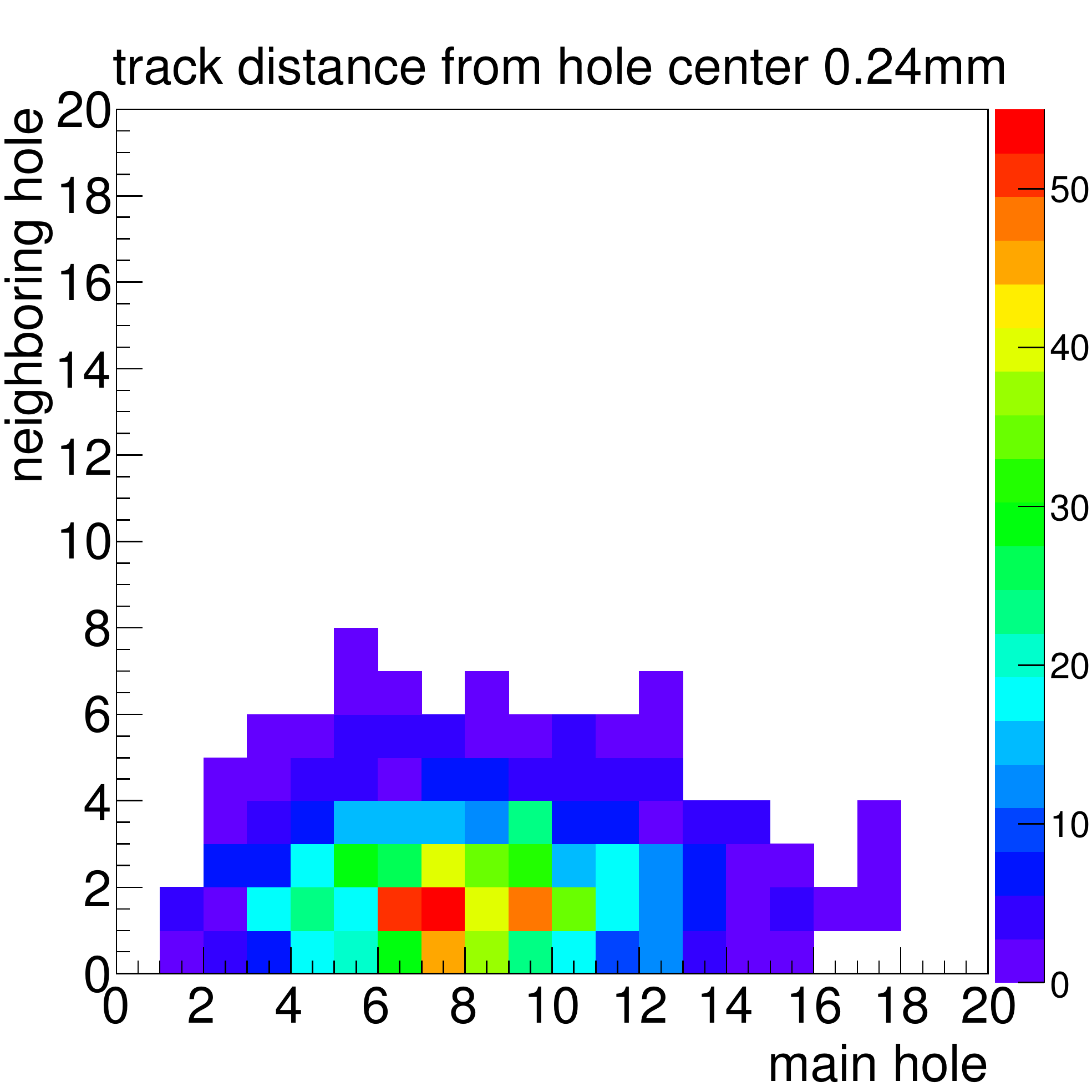}
\end{subfigure}
\begin{subfigure}[t]{0.3\textwidth}\caption{}
\includegraphics[scale=0.2]{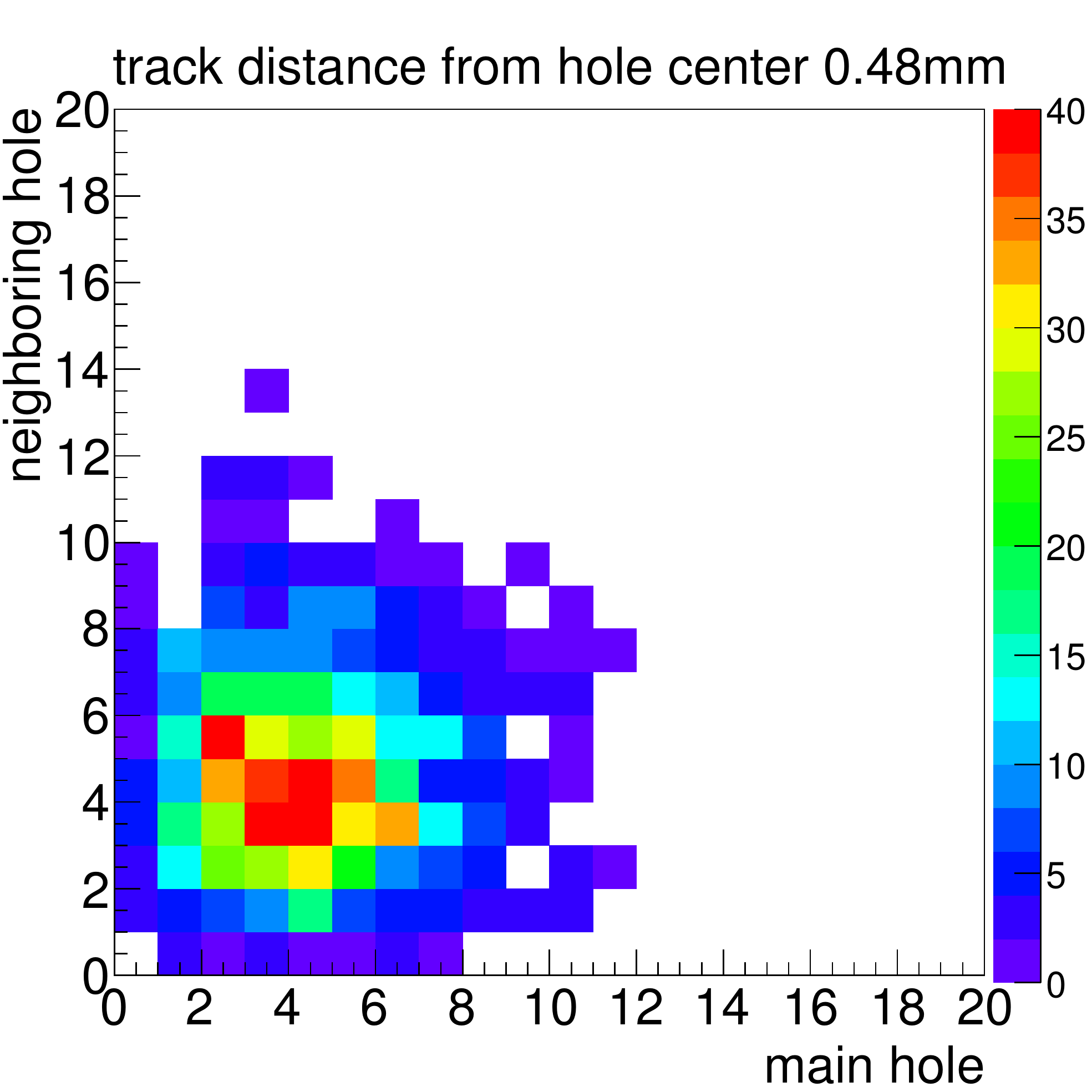}
\end{subfigure}\\
\begin{subfigure}[t]{0.3\textwidth}\caption{}
\includegraphics[scale=0.2]{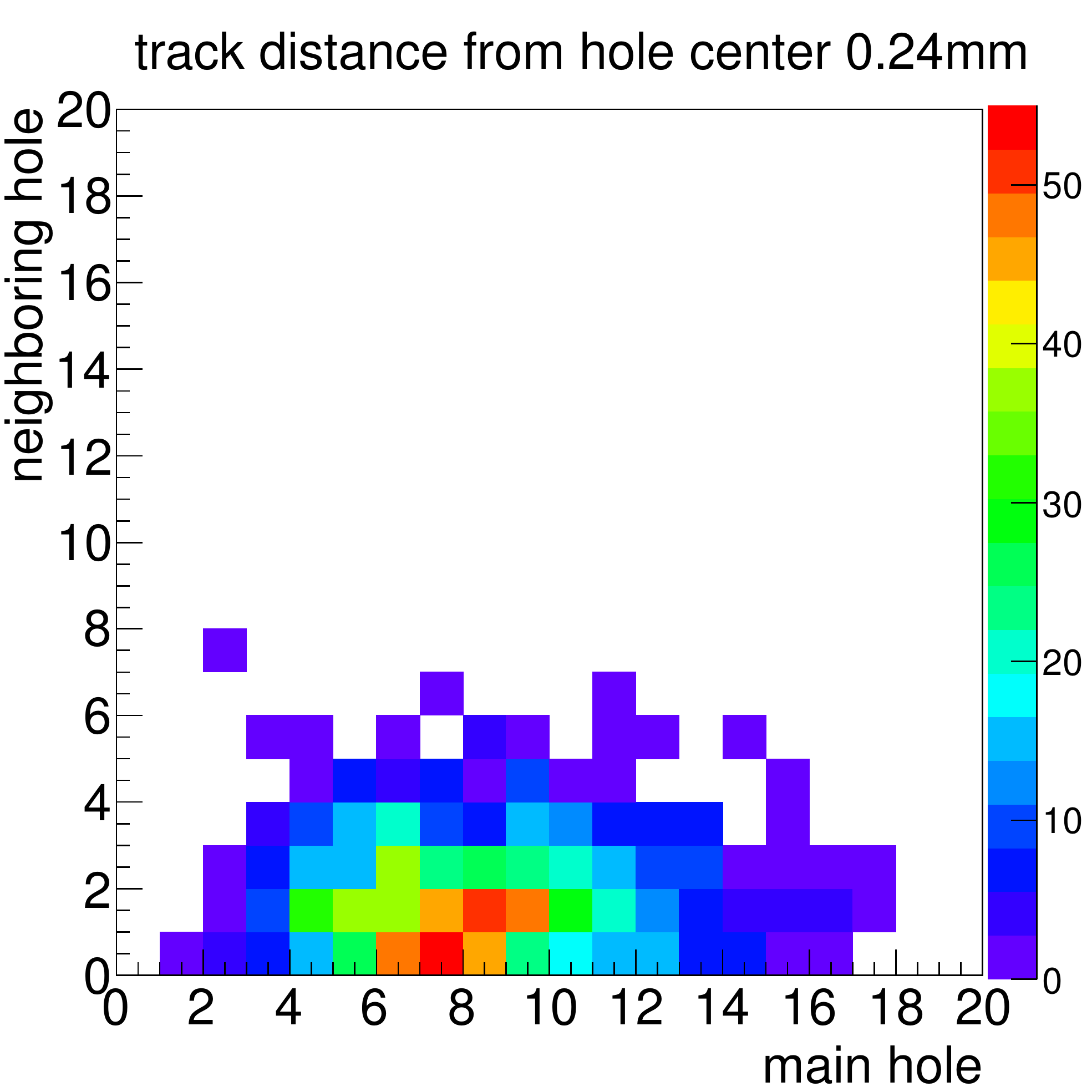}
\end{subfigure}
\begin{subfigure}[t]{0.3\textwidth}\caption{}
\includegraphics[scale=0.2]{figures/charge_sharing_2D_024mm.pdf}
\end{subfigure}
\begin{subfigure}[t]{0.3\textwidth}\caption{}
\includegraphics[scale=0.2]{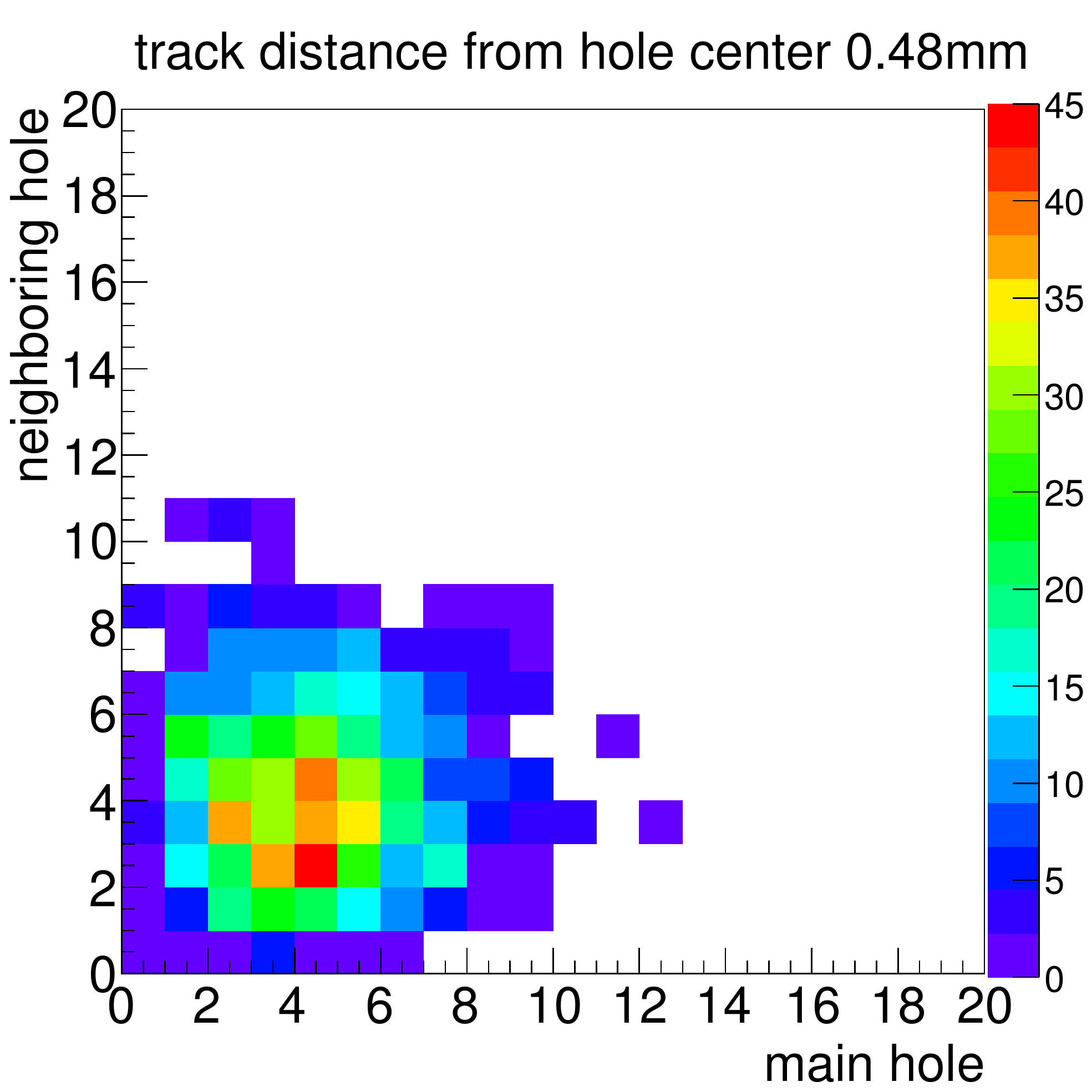}
\end{subfigure}
\caption{The fraction of primary charge reaching the closest hole and the first neighbor for perpendicular tracks at different distances from the hole center, before (a) and after (b) applying the time cut (see text).}\label{fig: charge sharing}
\end{figure}

As an input for the Monte Carlo simulation we created a library, which contains the information about the primary charges production and their drift into the THGEM holes for simulated muon track events. 
The steps that we followed for each event are:

\begin{enumerate}
\item \underline{Primary electrons production}: Heed~\cite{smirnov2005modeling} was used to generate clusters of ionization primary electrons (PE) along the 150~GeV muon track. 
\item \underline{Charge sharing and time-cut}: the Microscopic Drift Routine of Garfield~\cite{veenhof2015garfield} was used to drift the primary electrons along the field lines into the RPWELL holes. Considering the finite shaping time ($\sim$100~ns) of the APV25 readout chip~\cite{french2001design} and the arrival-time-spread of the primary electrons into the THGEM holes ($\sim$250~ns from-first-to-last~\cite{peisert1984drift}),  we decided to apply an event-dependent cut on the arrival time of the electrons before counting them. 
We let the electrons reaching the high field region to initiate avalanches. All avalanche electrons and ions were drifted to the RP top and THGEM top respectively. We used the Shockley Ramo theorem~\cite{ramo1939currents,shockley1938currents} to calculate the induced current on a uniform readout anode. This current pulse was then convoluted with the readout electronics transfer function~\cite{french2001design}. The resulting signal peaking time was considered as the time-cut. Only the primary electrons reaching a hole before then were considered. Figure~\ref{fig: charge sharing} shows, for perpendicular tracks at different distances from the hole-center, the fraction of primary charge reaching the closest hole and the first neighbor before and after applying the time cut. As can be seen, when the track is passing in the middle between two holes, the primary charge is shared equally.
\end{enumerate}

A set of 2000 events was produced for different track distances on the x-axis from a reference hole center (0 to 0.96~mm in steps of 0.05~mm). For each event we stored the number of primary electrons reaching the reference THGEM hole and also the next 7 holes in the direction of the track inclination. We neglected the holes in the other side since very rarely electrons were drifted there. Electrons reaching different holes along the y-axis are summed up.

A similar set was produced for different track incidence angles (0$^\circ$ to 40$^\circ$ in steps of 10$^\circ$). 

\subsection{Single-event Monte Carlo generator}
\label{sec: single event generator}

For each event the steps were: 
\begin{enumerate}
\item Generate randomly an x-position, x$\mathrm{_{tr}}$.
\item Draw randomly from the relevant primary-electrons library the number of PEs reaching the reference hole and the neighboring holes.
\item Calculate the total event charge and the fraction produced in each hole: 
\begin{enumerate}
\item Draw a random number g$\mathrm{_{ij}}$ from the single electron exponential gain distribution. This number represents the single primary electron avalanche gain, where \textit{i} corresponds to the hole where the primary electron ended and \textit{j} identifies the specific primary electron.  
\item Draw a random number Q distributed accordingly to the Landau fit to the cluster-charge spectrum (figure~\ref{fig: spectrum - residuals histo}-b).  This number represents the total charge produced in the event.
\item The fraction of the total charge produced in the hole \textit{i}, was then calculated as Q$\mathrm{_i}$= S$\mathrm{_i}\cdot$Q, where S$\mathrm{_i}$ = $\mathrm{\sum_j}$g$\mathrm{_{ij}}$/ $\mathrm{\sum_{ij}}$g$\mathrm{_{ij}}$.
\end{enumerate}
\item Produce the digitized signal (figure~\ref{fig: simulated signal}):
\begin{enumerate}
\item Create typical signal's double-Gaussian function f$\mathrm{_i}$ for each relevant hole. The mean of the distribution was set to the hole center with a small correction towards the track position (this effect is described in~\cite{rubin2013optical}).
\item Create histogram representing the readout. The bin width corresponds to the strips pitch, and the bin center corresponds to the strip position.
\item For each hole \textit{i}, fill the histogram with a number of events equal to Q$\mathrm{_i}$ distributed according to f$\mathrm{_i}$.
\end{enumerate}
\item Calculate the reconstructed event position: 
The reconstructed position of the simulated event was obtained by the charge-weighted centroid of the readout histogram x$\mathrm{_{rec}}$= $\mathrm{\sum x_iq_i}$/Q, where x$\mathrm{_i}$ is the bin center, and q$\mathrm{_i}$ is the bin content. The simulated residual for each event was calculated as RES= x$\mathrm{_{tr}}$-x$\mathrm{_{rec}}$. To effectively reproduce the effect of the ZSF threshold we considered only bins with >20 entries.
\end{enumerate}

\end{document}